\documentclass[fleqn,usenatbib]{mnras}

\usepackage[T1]{fontenc}
\usepackage{ae,aecompl}

\usepackage{graphicx}	
\usepackage{amsmath}	
\usepackage{amssymb}	
\usepackage{longtable}
\usepackage{newtxtext,newtxmath}

\newcommand{\msol}{M$_\odot \,$}
\newcommand{\rsol}{R$_\odot \,$}
\newcommand{\teff}{T$_{\rm eff}$}

\title[ZTF high-cadence Galactic Plane survey]{Year 1 of the ZTF high-cadence Galactic Plane Survey: Strategy, goals, and early results on new single-mode hot subdwarf B-star pulsators}

\author[T.Kupfer et al.]{
Thomas Kupfer,$^{1,2}$\thanks{E-mail: tkupfer@ttu.edu} 
Thomas A. Prince,$^{3}$
Jan van Roestel,$^{3}$
Eric C. Bellm,$^{4}$
\newauthor Lars Bildsten,$^{5,1}$
Michael W. Coughlin,$^{6}$
Andrew J. Drake,$^{3}$
Matthew J. Graham,$^{3}$
\newauthor Courtney Klein,$^{7}$
Shrinivas R. Kulkarni,$^{3}$
Frank J. Masci,$^{8}$
Richard Walters,$^{9}$
\newauthor Igor Andreoni,$^{3}$
Rahul Biswas,$^{10}$
Corey Bradshaw,$^{2}$
Dmitry A. Duev,$^{3}$
\newauthor Richard Dekany,$^{9}$
Joseph A. Guidry,$^{11}$
JJ Hermes,$^{12}$
Russ R. Laher,$^{8}$
\newauthor and Reed Riddle,$^{9}$
\\
$^{1}$Kavli Institute for Theoretical Physics, University of California, Santa Barbara, CA 93106, USA\\
$^{2}$Department of Physics and Astronomy, Texas Tech University, PO Box 41051, Lubbock, TX 79409, USA\\
$^{3}$Division of Physics, Mathematics and Astronomy, California Institute of Technology, Pasadena, CA 91125, USA\\
$^{4}$DIRAC Institute, Department of Astronomy, University of Washington, 3910 15th Avenue NE, Seattle, WA 98195, USA\\
$^{5}$Department of Physics, University of California, Santa Barbara, CA 93106, USA\\
$^{6}$School of Physics and Astronomy, University of Minnesota, Minneapolis, Minnesota 55455, USA\\
$^{7}$Department of Physics and Astronomy, University of California, Irvine, CA 92697, USA\\
$^{8}$IPAC, California Institute of Technology, 1200 E. California Blvd, Pasadena, CA 91125, USA\\
$^{9}$Caltech Optical Observatories, California Institute of Technology, Pasadena, CA 91125, USA\\
$^{10}$Department of Physics, The Oskar Klein Center, Stockholm University, AlbaNova, SE-10691 Stockholm, Sweden\\
$^{11}$Department of Astronomy, The University of Texas at Austin, Austin, TX 78712, USA\\
$^{12}$Department of Astronomy, Boston University, 725 Commonwealth Ave., Boston, MA 02215, USA
}

\date{Accepted XXX. Received YYY; in original form ZZZ}

\pubyear{2015}

\begin{document}
\label{firstpage}
\pagerange{\pageref{firstpage}--\pageref{lastpage}}
\maketitle

\begin{abstract}
We present the goals, strategy and first results of the high-cadence Galactic plane survey using the Zwicky Transient Facility (ZTF). The goal of the survey is to unveil the Galactic population of short-period variable stars, including short period binaries and stellar pulsators with periods less than a few hours. Between June 2018 and January 2019, we observed 64 ZTF fields resulting in $2990$\,deg$^2$ of high stellar density in ZTF-$r$ band along the Galactic Plane. Each field was observed continuously for 1.5 to 6\,hrs with a cadence of 40\,sec. Most fields have between $200$ and $400$ observations obtained over $2-3$ \,continuous nights. As part of this survey we extract a total of $\approx230$\,million individual objects with at least 80 epochs obtained during the high-cadence Galactic Plane survey reaching an average depth of ZTF-$r\approx20.5$\,mag. For four selected fields with 2\,million to 10\,million individual objects per field we calculate different variability statistics and find that $\approx$1-2\,\% of the objects are astrophysically variable over the observed period. We present a progress report on recent discoveries, including a new class of compact pulsators, the first members of a new class of Roche Lobe filling hot subdwarf binaries as well as new ultracompact double white dwarfs and flaring stars. Finally we present a sample of 12 new single-mode hot subdwarf B-star pulsators with pulsation amplitudes between ZTF-$r=20 - 76$\,mmag and pulsation periods between $P=5.8 - 16$\,min with a strong cluster of systems with  periods $\approx6$\,min. All of the data have now been released in either ZTF Data Release 3 or data release 4.
\end{abstract}




\section{Introduction}

Large scale optical time-domain surveys have opened a new window to study the variable sky providing hundreds to thousands of epochs across the whole sky. Starting with the Sloan Digital Sky Survey (SDSS; \citealt{yor00}), a new generation of wide-field optical surveys has exploited new affordable CCD detectors to open the frontier of data-intensive astronomy. This allows the study of stellar variability on different time-scales across the full magnitude range. In particular, surveys covering also low Galactic latitudes are well suited to study the Galactic distribution of photometric variable stars. Ground based surveys include the Optical Gravitational Lensing Experiment (OGLE; e.g. \citealt{sos15}), PTF \citep{law09}, the Vista Variables in the Via Lactea (VVV; \citealt{sai12}), ASAS-SN, \citep{sha14,jay18}) and most recently ATLAS \citep{ton18,hei18}. In particular the fastest stellar variabilities on timescales of minutes to hours are of great interest for a large number of scientific questions. This includes ultracompact binaries (UCBs), compact pulsators as well as fast flaring stars. 

UCBs are a class of binary stars with orbital periods less than about 60min, consisting of a neutron star (NS)/white dwarf (WD) primary and a Helium-star (He-star)/WD/NS secondary. These UCBs are sources of low-frequency gravitational wave (GW) signals as probed by the Laser Interferometer Space Antenna (\emph{LISA}) and are crucial to our understanding of compact binary evolution and offer pathways towards Type Ia and other thermonuclear supernovae. Systems with orbital periods $<$20min will be the strongest Galactic \emph{LISA} sources and will be detected by \emph{LISA} within weeks after its operation begins \citep{nel04, str06, nis12, lit13, kor17,kre17, kup18, lam19, bur19, bur20} and as such are ideal multi-messenger sources \citep{sha12, sha14a, lit18, bak18, kup19a}.

Short time-scale photometric variations can also originate from astrophysical changes within the internal structure or atmosphere of the star. Such sources are flaring stars, pulsating stars or white dwarfs. Prominent examples of rapidly pulsating stars are $\delta$ Scuti stars \citep{bre00}, SX Phoenicis variables \citep{nem90} or Ap or Am stars \citep{ren91}, and pulsating white dwarfs (ZZ Ceti variables; \citealt{fon08}) as well as pulsating extremely low mass white dwarfs \citep{her13}. The stars exhibit pulsation amplitudes of less than one percent up to several tens of percent on timescales of few to tens of minutes. 

Recently, a new class of short period pulsating hot stars known as Blue Large-Amplitude Pulsators (BLAPs) was discovered by \citet{pie17}. Their pulsation periods are typically between 20-40 minutes. \citet{rom18} and \citet{byr18} proposed that the BLAPs are hot pre-helium white dwarfs that are cooling and contracting, with masses in the range 0.3-0.35\msol. A recent study by \citet{men20} suggests that BLAPs could be the surviving companions of type\,Ia supernovae. In this scenario mass transfer from the BLAP progenitor leads to the detonation of a white dwarf companion, leaving behind a BLAP as single star. Their pulsation properties can best be explained by fundamental radial mode pulsators. However, the known sample of BLAPs is small and in-homogeneous, discovered only in the OGLE survey.

\begin{table}
	\centering
	\caption{Overview of the high-cadence Galactic Plane survey}
	\label{tab:survey_overview}
	\begin{tabular}{lccr} 
		\hline
		Period & \#fields  & sky coverage & Filter \\
               &           &    (deg$^2$)    &  \\
		\hline
		06-15-2018 - 07-31-2018 & $16$ & $750$ & ZTF-$r$ \\
		08-03-2018 - 08-18-2018 & $14$ & $650$ & ZTF-$r$ \\
		11-15-2018 - 01-15-2019 & $34$ & $1590$ & ZTF-$r$ \\
		\hline
	\end{tabular}
\end{table}



A number of fast cadence ground-based surveys have been executed to study the variable sky down to a few minutes period. The main goal of these high-cadence surveys is the discovery of rapid brightness variations seen in UCBs, compact pulsators as well as fast flaring stars. The first survey at low Galactic latitudes targeting short-period systems was the Rapid Temporal Survey (RATS; \citealt{ram05,bar11}) covering a total of 46\,deg$^2$. Another more recent survey is the OmegaWhite (OW) survey, which covers a total of 400\,deg$^2$ at low Galactic latitudes ($|b|<$\,10$^\circ$) as well as in the Galactic Bulge using high-cadence optical observations. Two neighboring one square degree fields are alternatingly observed in 39 second exposures over an observing duration of 2 hours, with an observational median cadence of $\approx2.7$ minutes per field \citep{mac15,tom16,mac17,mac17a,kup17a}.

\begin{figure*}
\centering
\centering
\includegraphics[width=\textwidth]{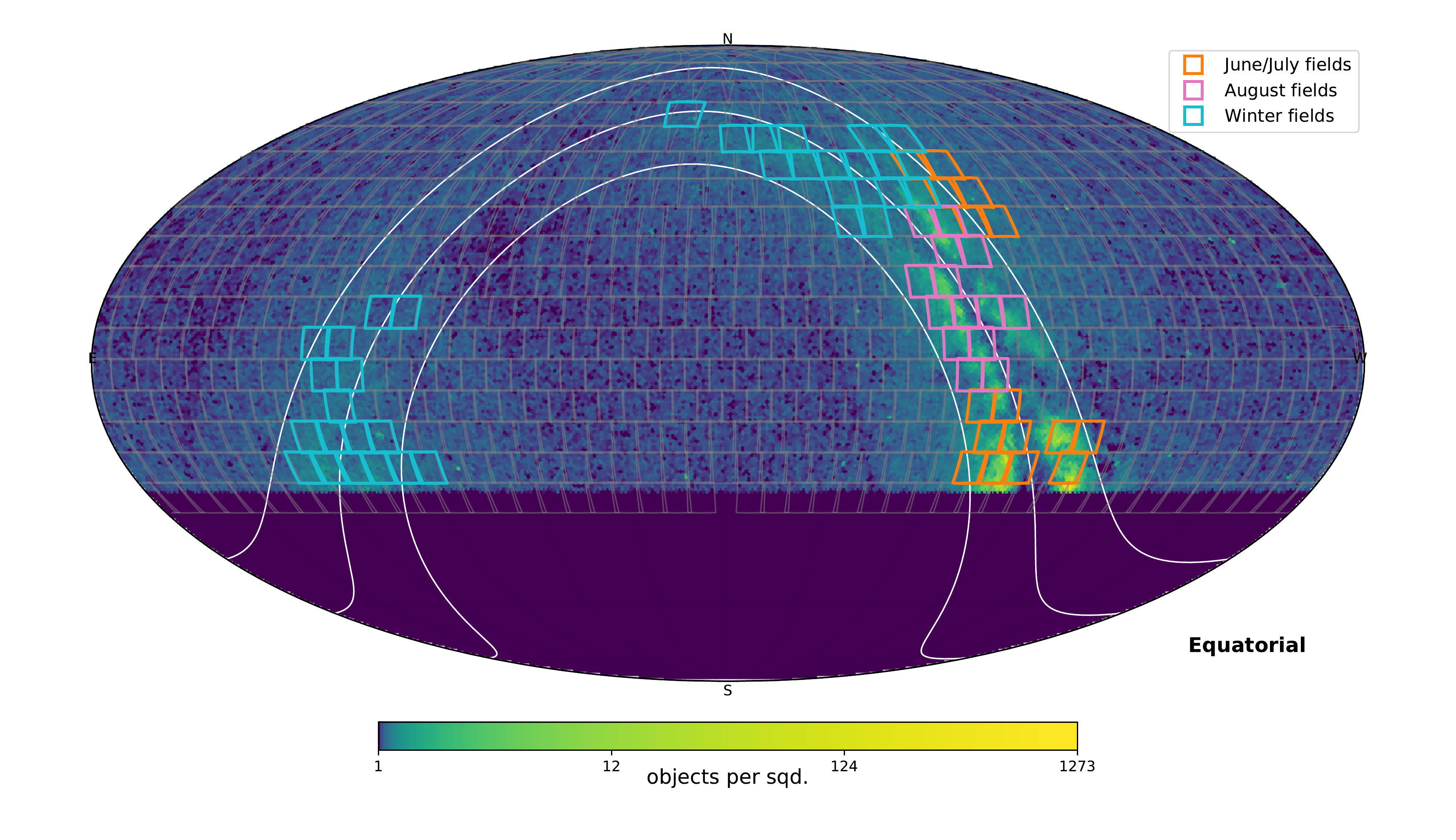} 	\vspace{-1cm}
    \caption{Sky density of candidate white dwarfs and hot subdwarfs selected from Gaia DR2 with the selected fields of the ZTF high-cadence Galactic Plane survey observed in ZTF year-1. The squares show individual ZTF fields which have been observed in high-cadence Galactic Plane observations. The white lines correspond to the Galactic equator and $\vert b\vert=15$\,deg}
   \label{fig:ztf_fields}
\end{figure*}

\begin{figure}
\centering
\centering
\includegraphics[width=0.48\textwidth]{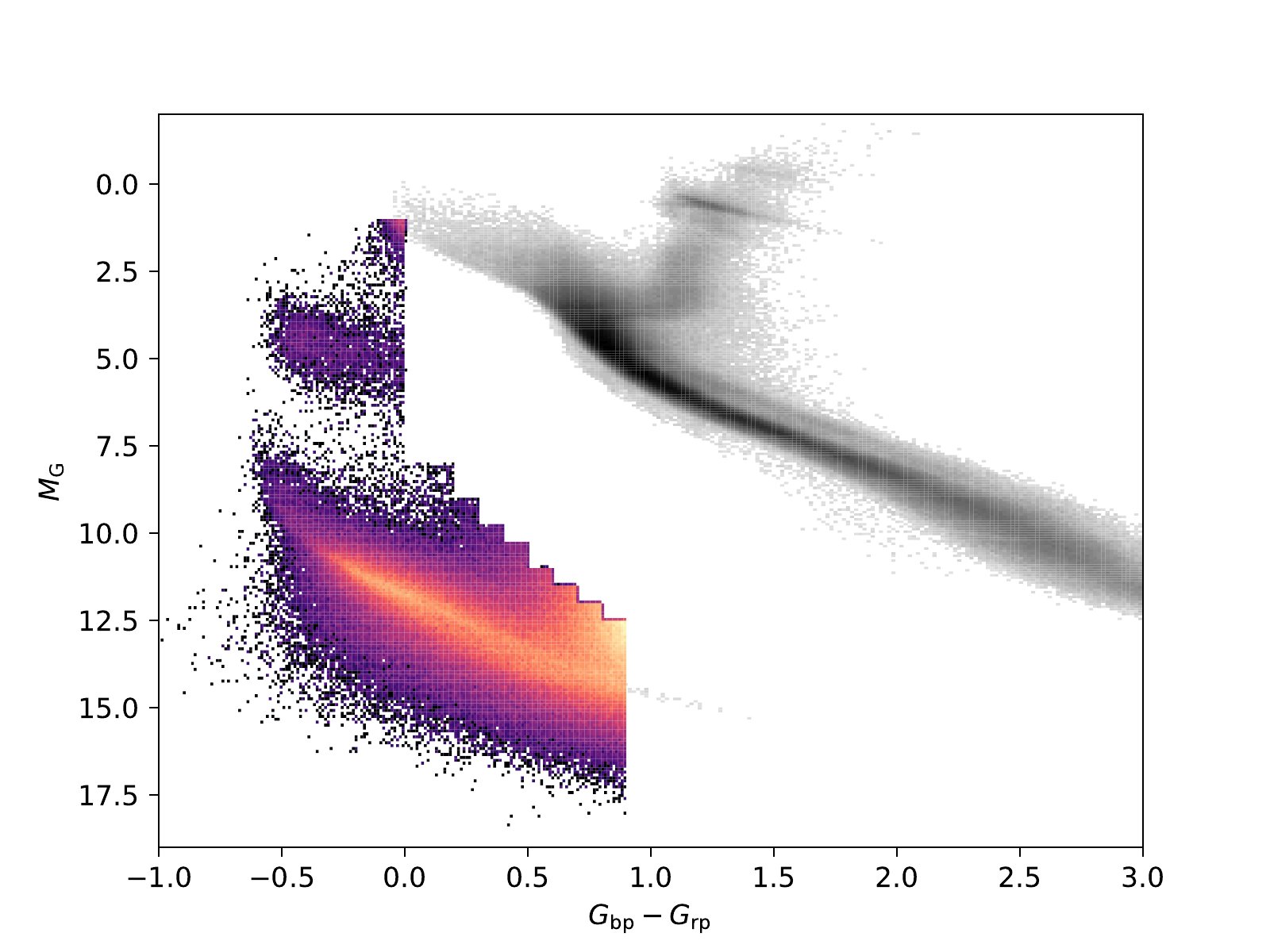} 
    \caption{Hertzsprung Russell diagram of the objects used for the field selection. The grey shaded region corresponds to the underlying Hertzsprung Russell diagram showing the position of the main sequence and the red giant branch. The color-coded region corresponds to the objects which were selected for the field selection. The clump around $M_{\rm G}\approx5$ corresponds to the hot subdwarfs whereas the region below corresponds to the white dwarfs.}
   \label{fig:ztf_hr}
\end{figure}

As part of the Zwicky Transient Facility (ZTF), the Palomar 48-inch (P48) telescope images the sky every clear night conducting several surveys including a Northern Sky Survey with a 3-day cadence, as well as smaller surveys such as a 1-day Galactic Plane survey and simultaneous observations of the Northern TESS sectors \citep{gra19,bel19,bel19a,roe19}. As part of the partnership share of ZTF we conducted a dedicated high-cadence Galactic Plane survey with a cadence of 40\,sec at low Galactic latitudes aiming to find ultracompact binaries and compact pulsators. During that dedicated survey we either observed one field or alternated between two adjacent fields continuously for $\approx$1.5-3\,hours on two to three consecutive nights in the ZTF-$r$ band. Here we present an overview of the ZTF high-cadence Galactic Plane survey executed in ZTF year 1. We present the observing strategy as well as some survey statistics. We show a progress report and finalize with some early results from the survey. In Section\,\ref{sec:design} we discuss the observing strategy as well as the field selection and the data processing. In Section\,\ref{sec:fields} we present results for four representative fields and in Section\,\ref{sec:progress} we give a progress report of already published results. In Section\,\ref{sec:early} we show some new early results from the high-cadence Galactic Plane survey and summarize and conclude in Section\,\ref{sec:sum}.

\begin{figure*}
\centering
\centering
\includegraphics[width=\textwidth]{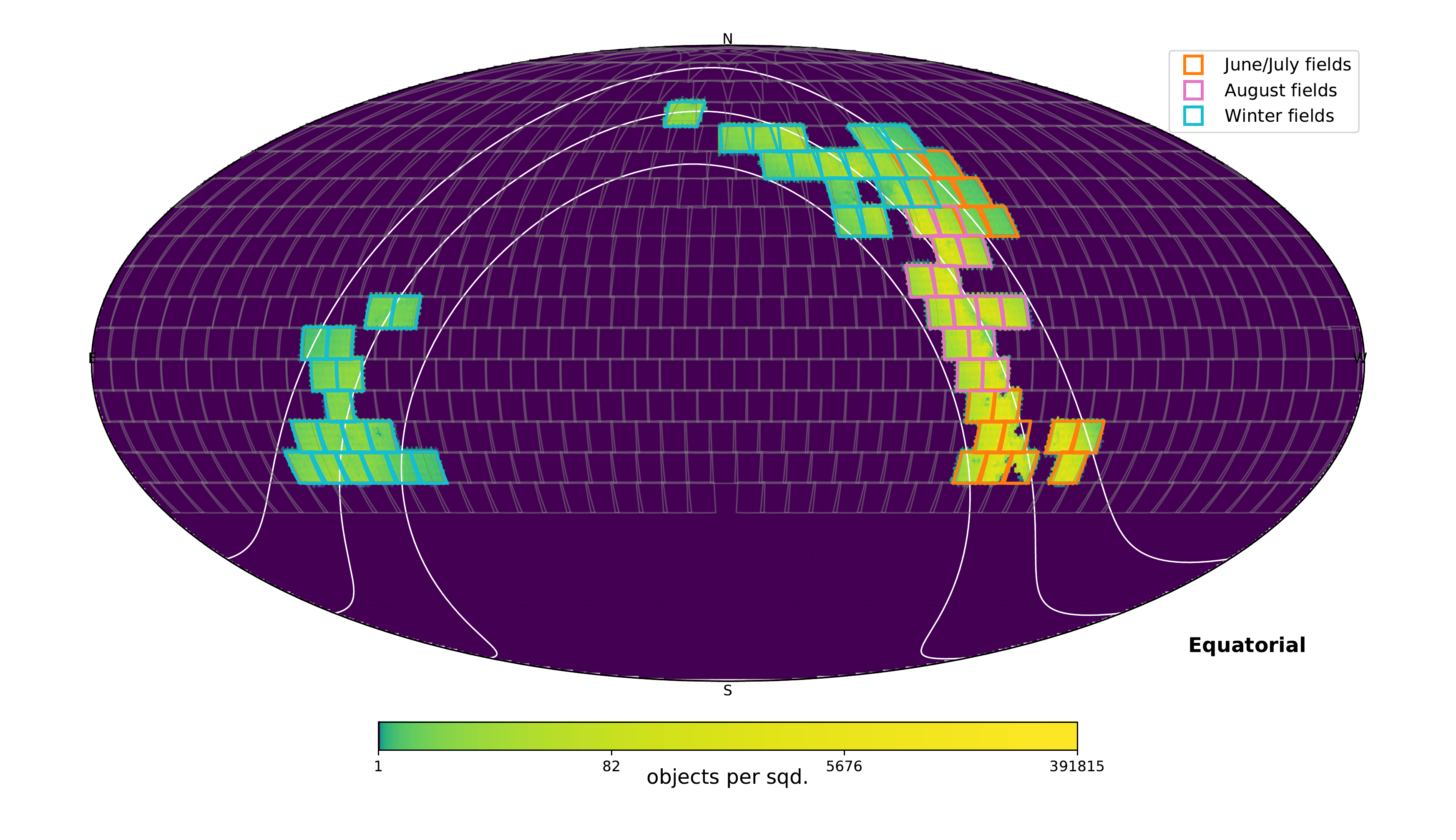} 	\vspace{-1cm}
    \caption{Sky density of individual objects with at least 80 epochs from the ZTF high-cadence Galactic Plane survey observed in ZTF year-1. The squares show individual ZTF fields which have been observed in high-cadence Galactic Plane observations. The white lines correspond to the Galactic equator and $\vert b\vert=15$\,deg}
   \label{fig:ztf_selected}
\end{figure*}

\section{Design of the high-cadence Galactic Plane survey}\label{sec:design}
ZTF uses a 47\,deg$^2$ camera consisting of 16 individual CCDs each 6k$\times$6k covering the full focal plane of the P48 telescope. The ZTF high-cadence Galactic Plane survey covered a total of $\approx$2990\,deg$^2$ split in 64 individual ZTF fields observed over three observing blocks; mid-June to July 2018, two weeks in August 2018 and November 15th 2018 to January 15th 2019. All observations of the high-cadence Galactic Plane survey were obtained in ZTF-$r$ band. Each fields has $\approx200-400$ epochs. The images were processed using ZTF data processing pipeline described in full detail in \citet{mas19}. 

During June and July we observed 16 fields covering $\approx750$\,deg$^2$ mostly at low Galactic longitudes. In June we observed every field continuously for 1hr15min and in July for 1hr25min. All fields were observed over $2$ or $3$ consecutive nights. In August, 14 fields covering $\approx650$\,deg$^2$ were observed over two weeks (see Tab.\,\ref{tab:survey_overview}). We alternated between two adjacent fields continuously for 2hrs 40min each night. The same fields were repeated the following night. The observations in June/July and August were done under stable conditions with an average seeing of $\approx2$\,arcsec. We lost only a total of 5 nights due to weather during June/July and August observations. 

Between 2018 November 15 and 2019 January 15 high-cadence Galactic Plane observations were scheduled for every night. We observed an additional 33 fields covering $\approx1550$\,deg$^2$ with stable weather conditions (see Tab.\,\ref{tab:survey_overview}). The overall strategy varied during those two months due to unstable weather. Most fields were observed alternating with adjacent fields but in particular low declination fields were observed continuously. Because more time was available each night most fields were observed for $\approx$3\,hrs. However, due to the unstable weather conditions about half of the fields were only observed in a single night and not repeated in subsequent nights. All other fields were observed over $2$ or $3$ nights. Although the seeing varies strongly between 1.7\,arcsec and 4\,arcsec, each field has a limiting magnitude of $>19.5$\,mag. A detailed overview of the fields observed in stable weather conditions is given in the Appendix in Table\,\ref{tab:fields_summer}, \ref{tab:fields_fall} and \ref{tab:fields_winter}.

\subsection{Field selection}
The main science driver for the survey is to find and study UCBs consisting of fully degenerate or semi-degenerate stars. Hence we selected ZTF fields based on the density of objects residing well below the main-sequence, including mostly white dwarfs and hot subdwarf stars. 

To achieve this, we extracted objects with absolute magnitudes which placed them below the main-sequence based on Gaia data release 2 \citep{gai16,gai18}. Only objects with declinations $>-30^\circ$ were selected. As the goal was not to extract a clean sample we used a very relaxed tolerance for the parallax precision ($\varpi/\sigma_{\varpi}>3$) and did not include any additional quality cuts, resulting in $\approx$350,000 individual objects (Fig.\,\ref{fig:ztf_hr}). For each object the ZTF field was calculated and the fields with the largest number of individual objects were selected for our survey. Selected fields with the highest density have $\approx1000$ selected objects per deg$^2$. The final field selection was mainly driven by stellar density but also included other aspects like visibility and weather. Fig.\,\ref{fig:ztf_fields} shows the sky density of the selected sources overplotted with the fields observed as part of the high-cadence Galactic Plane survey. As the stellar density is highest at low Galactic latitudes most fields are located at Galactic latitudes $\leq15$\,deg. 

\begin{figure*}
\centering
\centering
\includegraphics[width=0.47\textwidth]{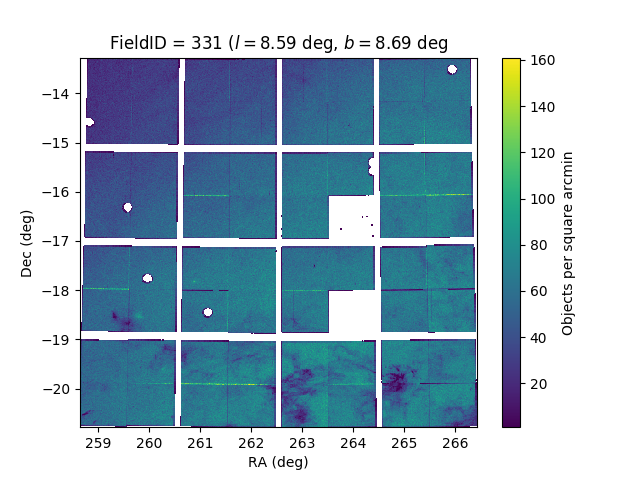} 
\includegraphics[width=0.47\textwidth]{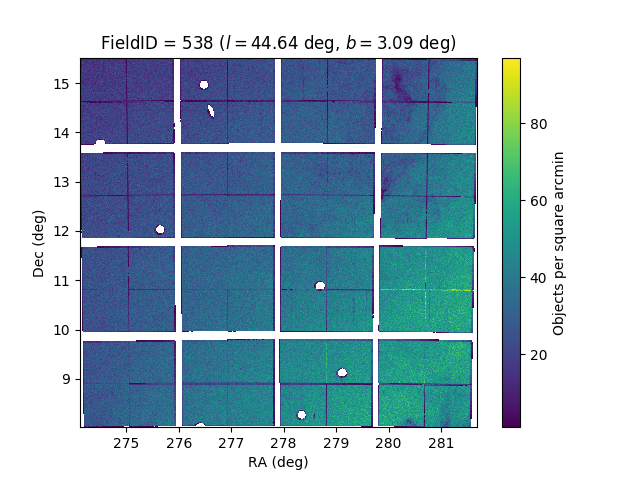} 
\includegraphics[width=0.47\textwidth]{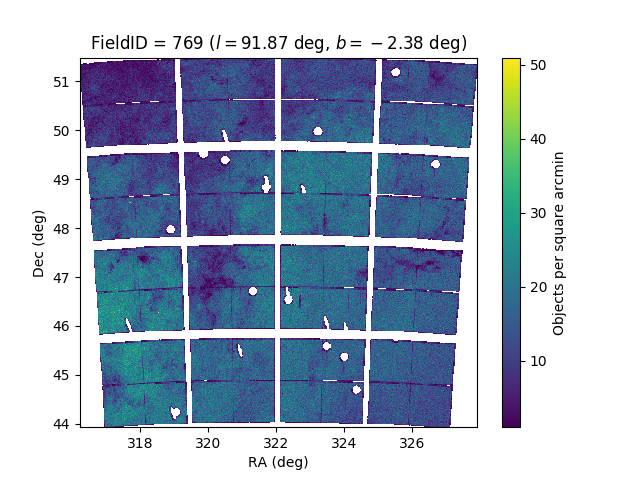} 
\includegraphics[width=0.47\textwidth]{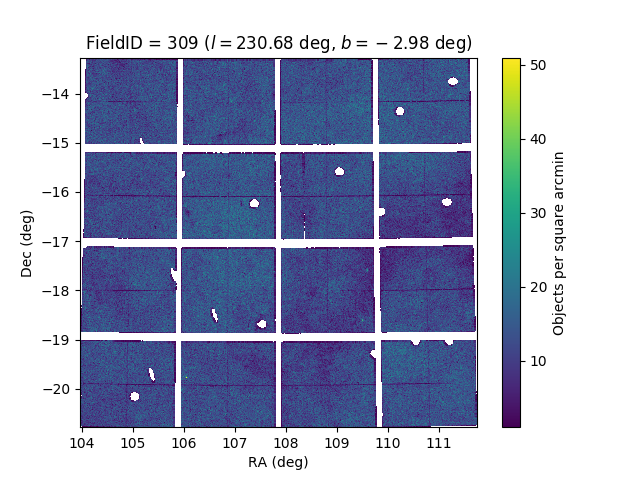} 
    \caption{Sky density of ZTF objects with at least 80 epochs obtained during the high-cadence Galactic Plane survey for four selected fields. The white circles are masked regions due to saturated stars and the horizontal and vertical gaps are chip gaps between the 16 CCDs. The two white square areas in FieldID 331 are individual quadrants which were not processed.}
   \label{fig:ztf_skydens}
\end{figure*}

\subsection{Data processing and light curve extraction}
Data processing and light curve generation follows the standard procedure for ZTF and occurs at the Infrared Processing and Analysis Center, Caltech. The raw camera image data is first instrumentally calibrated and astrometric solutions are derived using the Gaia DR1 catalog. Sources are detected and fluxes measured using both aperture \citep{ber96} and PSF-fit photometry \citep{ste87}. Sources are photometrically calibrated using the Pan-STARRS1 DR1 catalog. The epochal image data are then co-added within their respective survey fields and camera readout channels to construct reference images. Each reference image is constructed using a minimum of 15 and maximum of 40 good quality epochal images, yielding depths of $\approx2-2.5$\,mag deeper than the single-epoch images. 

The reference image co-adds are archived for use in other downstream processing: image differencing and light curve construction. To support light curve generation, sources are first detected and extracted from each reference image using PSF-fit photometry by running the DAOPhot utility \citep{ste87}. These sources provide the seeds to facilitate  positional  cross-matching  across  all  the single-epoch-based  PSF  extraction  catalogs  going  back  to the beginning of the survey. PSF-fitting on the single epoch images is also performed using DAOPhot. These single epoch images are direct single exposure images, not difference images. A fixed source-match radius of 1.5 arcsec is used for the positional matching. Only the PSF-fit-derived positions and photometry, along with selected PSF-fit metrics are retained during the  source-matching process. Aperture photometry is not propagated to the light curve metadata.


Further details of the data processing, PSF-fit photometry software, light curve generation, formats, and overall performance on photometric accuracy are described in \citet{mas19}. We extracted the light curves from each object which has at least $80$ epochs obtained as part of the high-cadence Galactic Plane survey from the ZTF light curve database ``Kowalski'' and extracted a total of $\approx230$\,million light curves across all observed fields. Figure\,\ref{fig:ztf_selected} shows the sky density of the extracted objects overplotted with the fields observed as part of the high-cadence Galactic Plane survey.




\begin{figure*}
\centering
\centering
\includegraphics[width=0.49\textwidth]{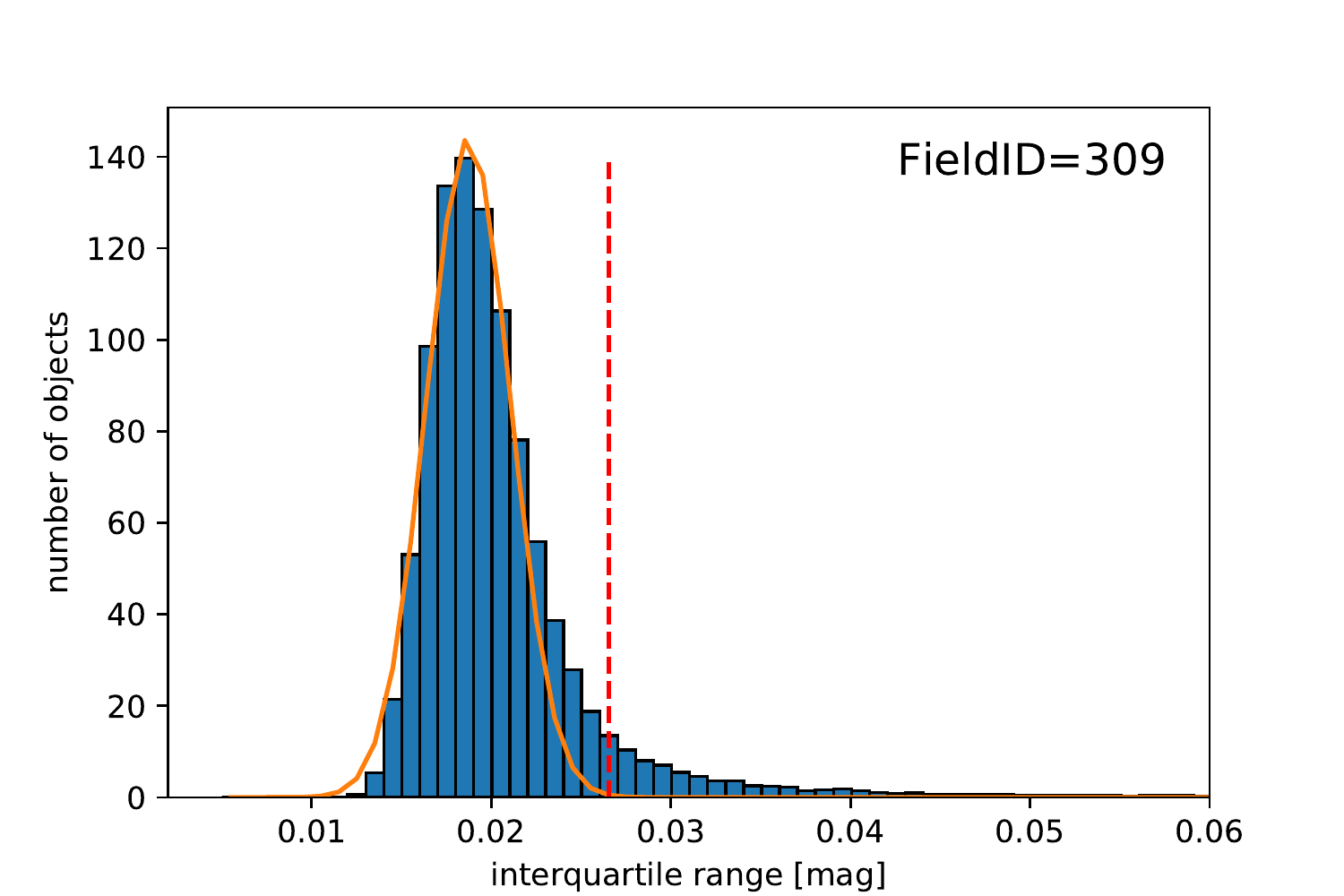} 
\includegraphics[width=0.49\textwidth]{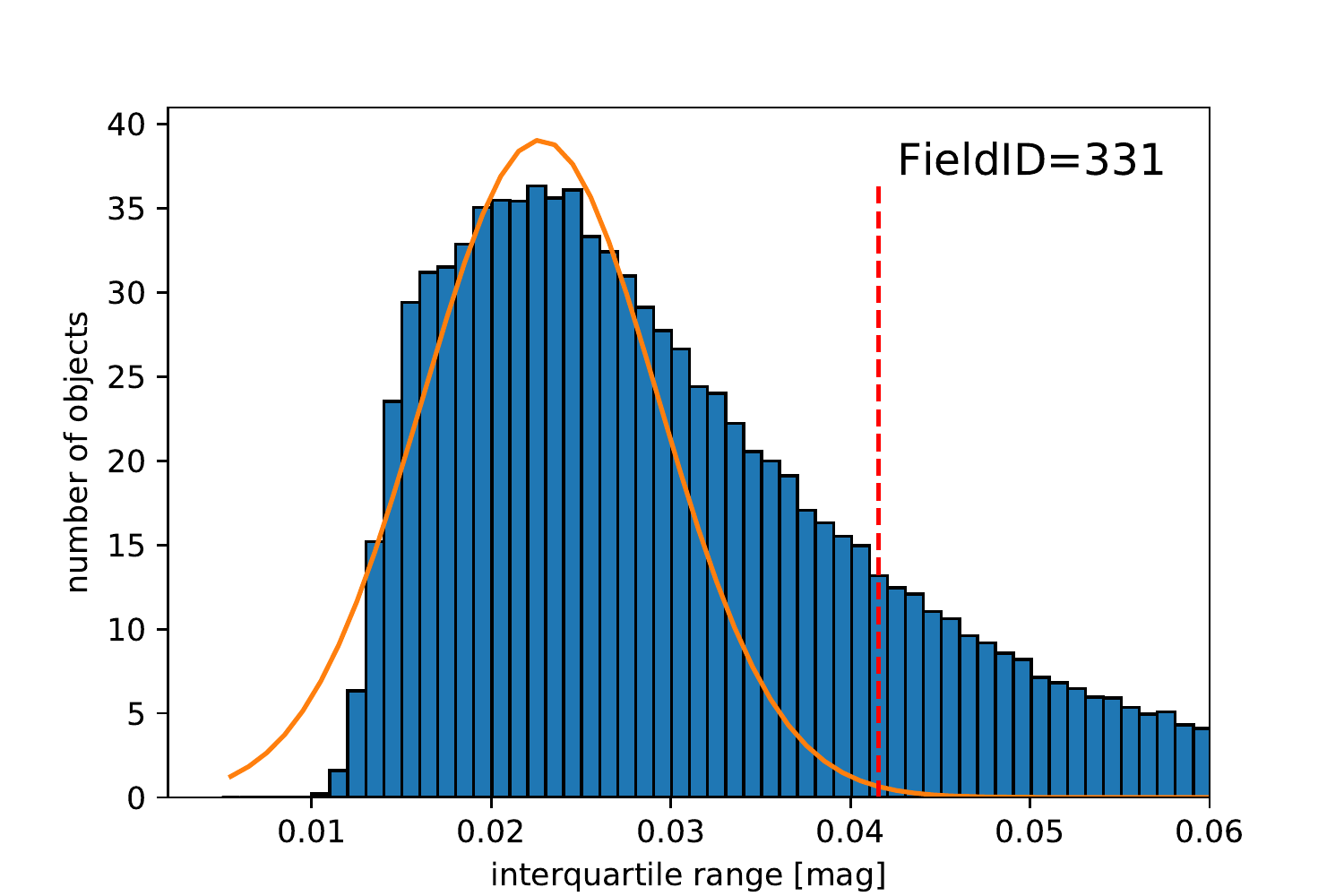} 
    \caption{Histogram of the interquartile range extracted at a ZTF-$r$ median magnitude between 15.5 and 16 with the best Gaussian fit and the limit above which an object is called variable. The fit to the histogram in FieldID=331 is poor which is likely due to a large number of blended sources. See Sec.\,\ref{sec:fields} for more details.}
   \label{fig:ztf_histo}
\end{figure*}

\begin{figure*}
\centering
\centering
\includegraphics[width=0.49\textwidth]{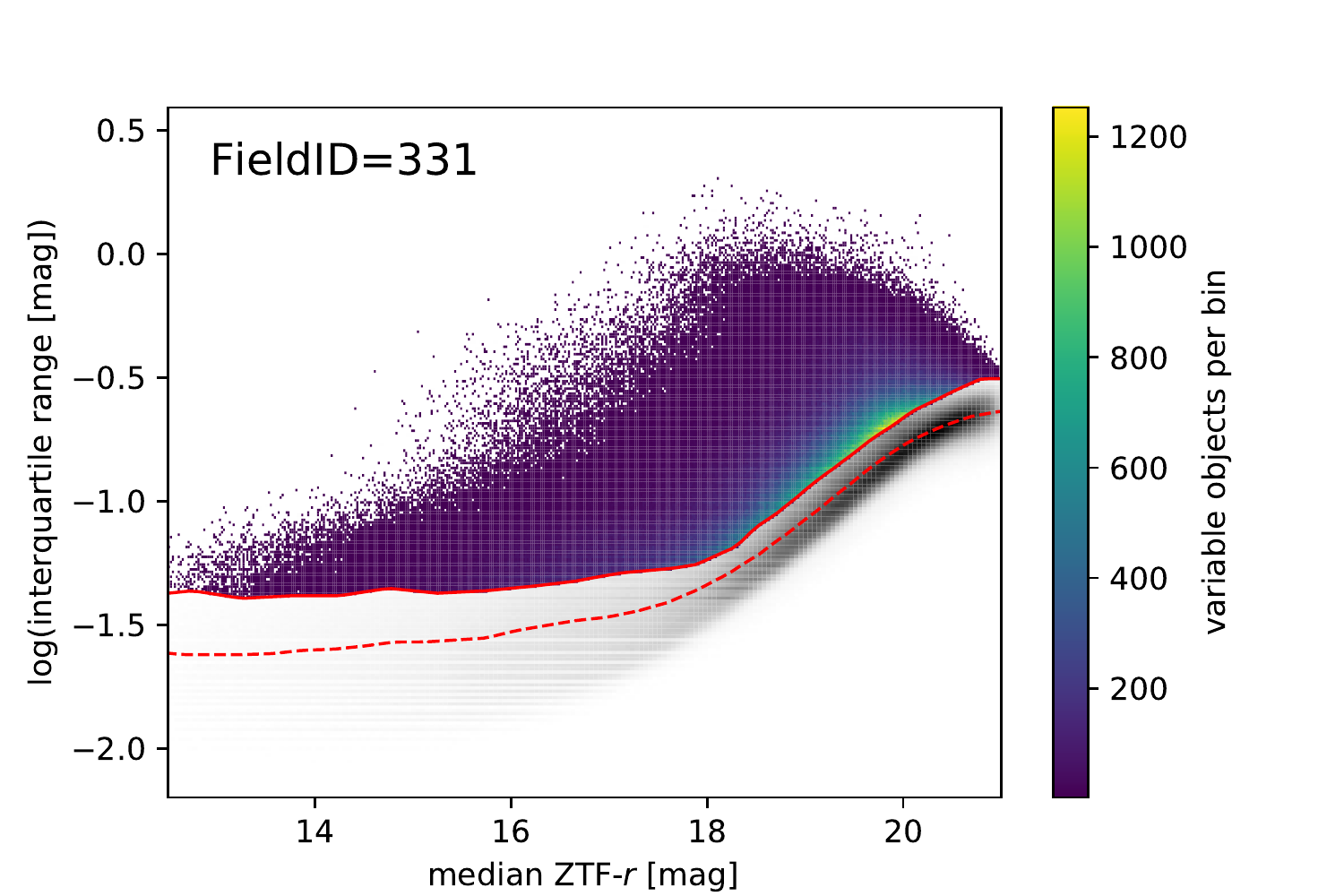} 
\includegraphics[width=0.49\textwidth]{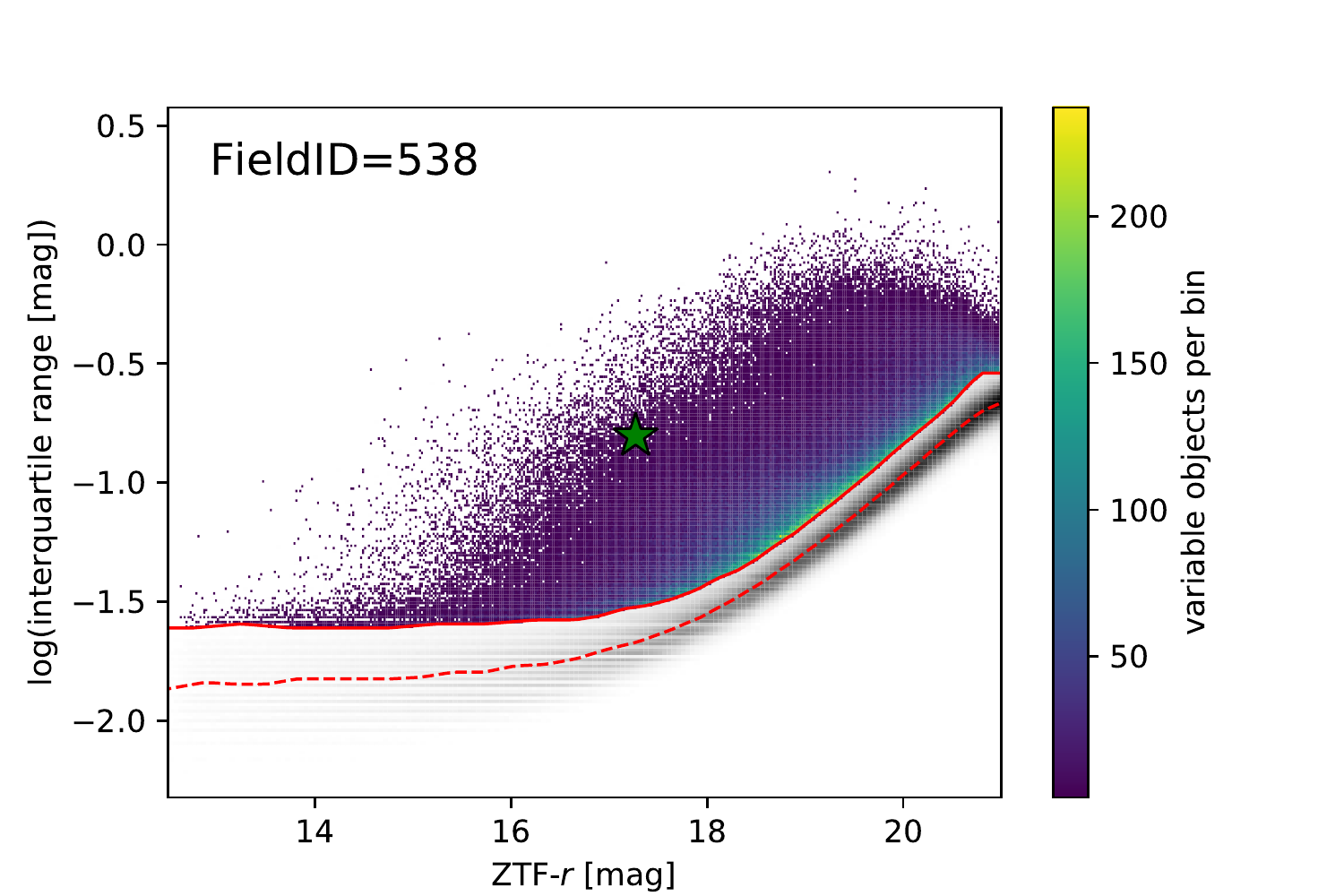} 
\includegraphics[width=0.49\textwidth]{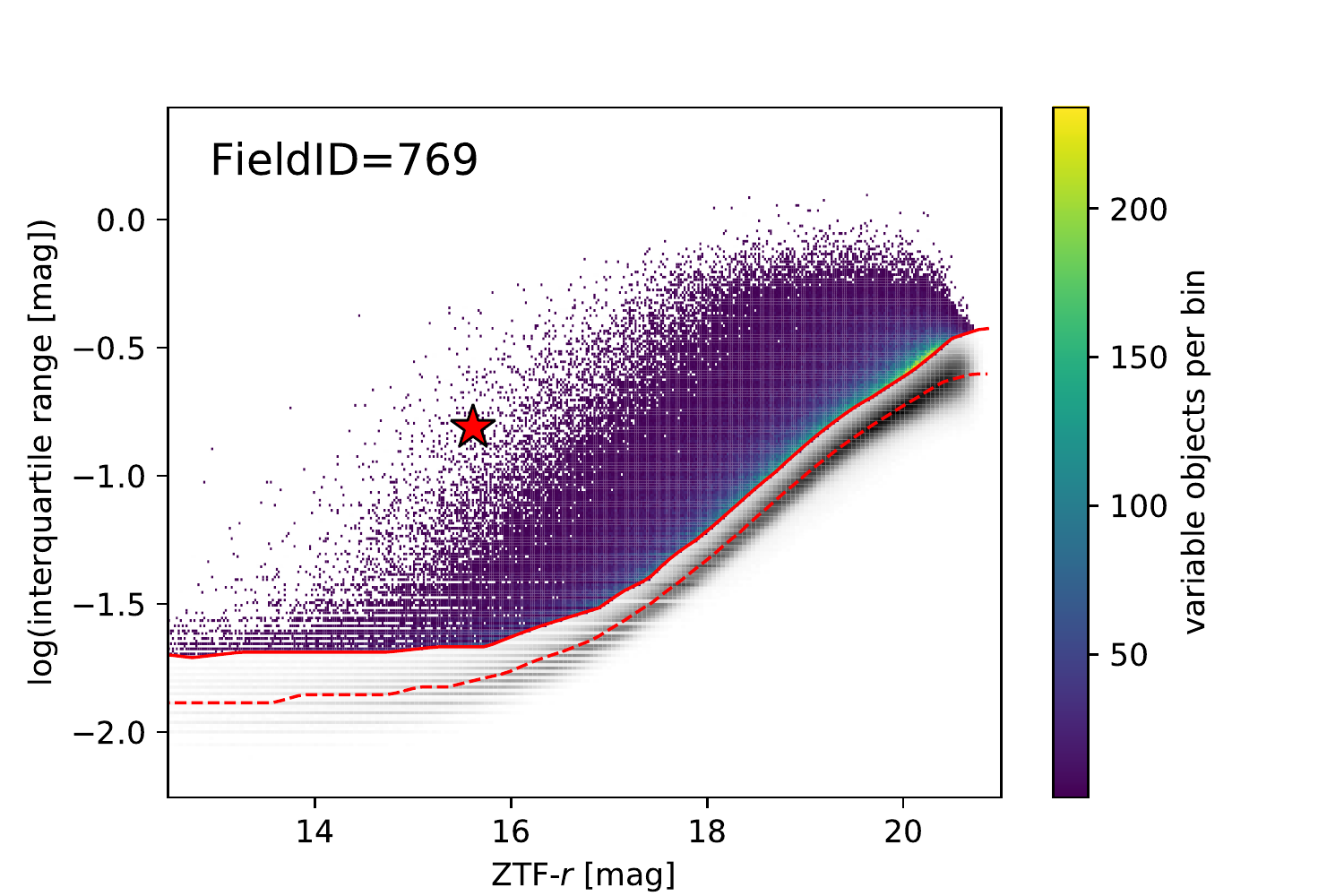} 
\includegraphics[width=0.49\textwidth]{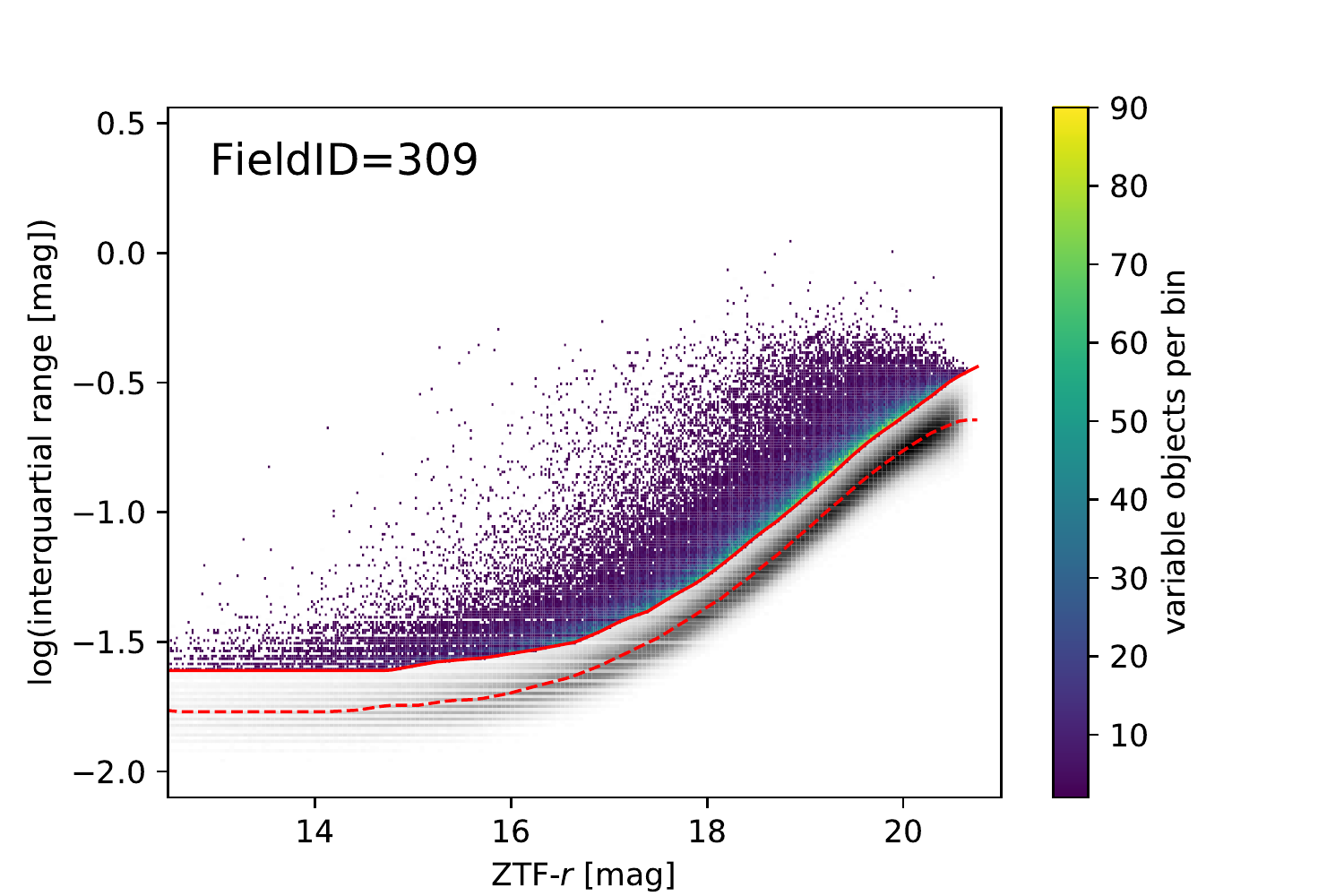} 
    \caption{Interquartile range for each field. The grey shaded region corresponds to the underlying distribution of all objects in that field. The color-coded region corresponds to the objects which are possibly variable. The dashed red line shows the robust median of the underlying grey shaded region. The solid red line corresponds to the limit for potentially variable objects. The red star marks the position of ZTF\,J2130 \citep{kup20} and the green star marks the position of high-gravity BLAP-3 \citep{kup19}.}
   \label{fig:ztf_variab}
\end{figure*}

\section{Results from representative fields}\label{sec:fields}
We select four representative fields with different stellar densities and calculate statistics for each individual light curve, including median magnitude, reduced $\chi^2$, interquartile range, skewness, inverse von-Neumann, Stetson J and Stetson K statistics. We refer to Table 2 in \citet{cou20} for a detailed description of the individual statistics. These statistics are used to evaluate light curve variability and identify variable objects in the high-cadence Galactic Plane data. 

The selected fields are FieldID=331 ($l=8.59$\,deg, $b=8.69$\,deg), FieldID=538 ($l=44.64$\,deg, $b=3.09$\,deg), FieldID=769 ($l=91.87$\,deg, $b=-2.38$\,deg) and FieldID=309 ($l=230.68$\,deg, $b=-2.98$\,deg). Figure\,\ref{fig:ztf_skydens} shows the stellar density of the four selected fields. The different coloring correspond to different densities. Some structure of lower and higher density regions can be seen in the fields. The white circles are masked regions due to saturated stars and the horizontal and vertical gaps are chip gaps between the 16 CCDs. The two white square areas in FieldID 331 are individual quadrants which were not processed. FieldID=331 was observed in three consecutive nights for 1.5\,hrs each night. FieldID=538 was observed for two consecutive nights for 1.5\,hrs and 2.5\,hrs. FieldID=769 was observed for two consecutive nights for 3\,hrs per night and FieldID=309 was observed for two consecutive nights for 1.7\,hrs and 3\,hrs. See Table\,\ref{tab:fields_summer}, \ref{tab:fields_fall} and \ref{tab:fields_winter} for more details. The four selected fields represent a large range of stellar densities from $\approx10$\,million individual light curves for FieldID=331, $\approx5.8$\,million for FieldID=538, $\approx3.8$\,million for FieldID=769 and $\approx2.3$\,million for FieldID=309.

\begin{figure*}
\centering
\includegraphics[width=0.49\textwidth]{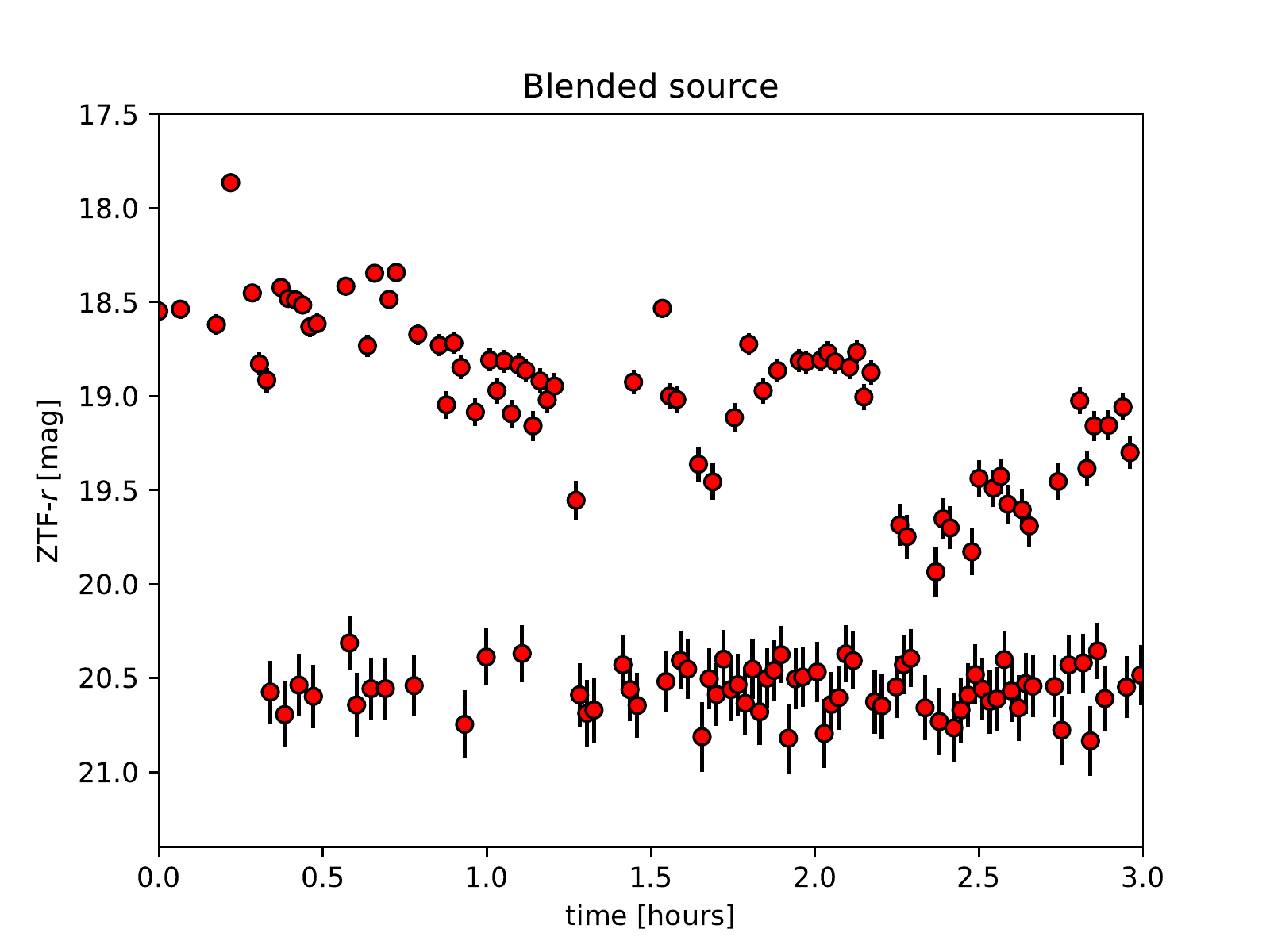} 
\includegraphics[width=0.49\textwidth]{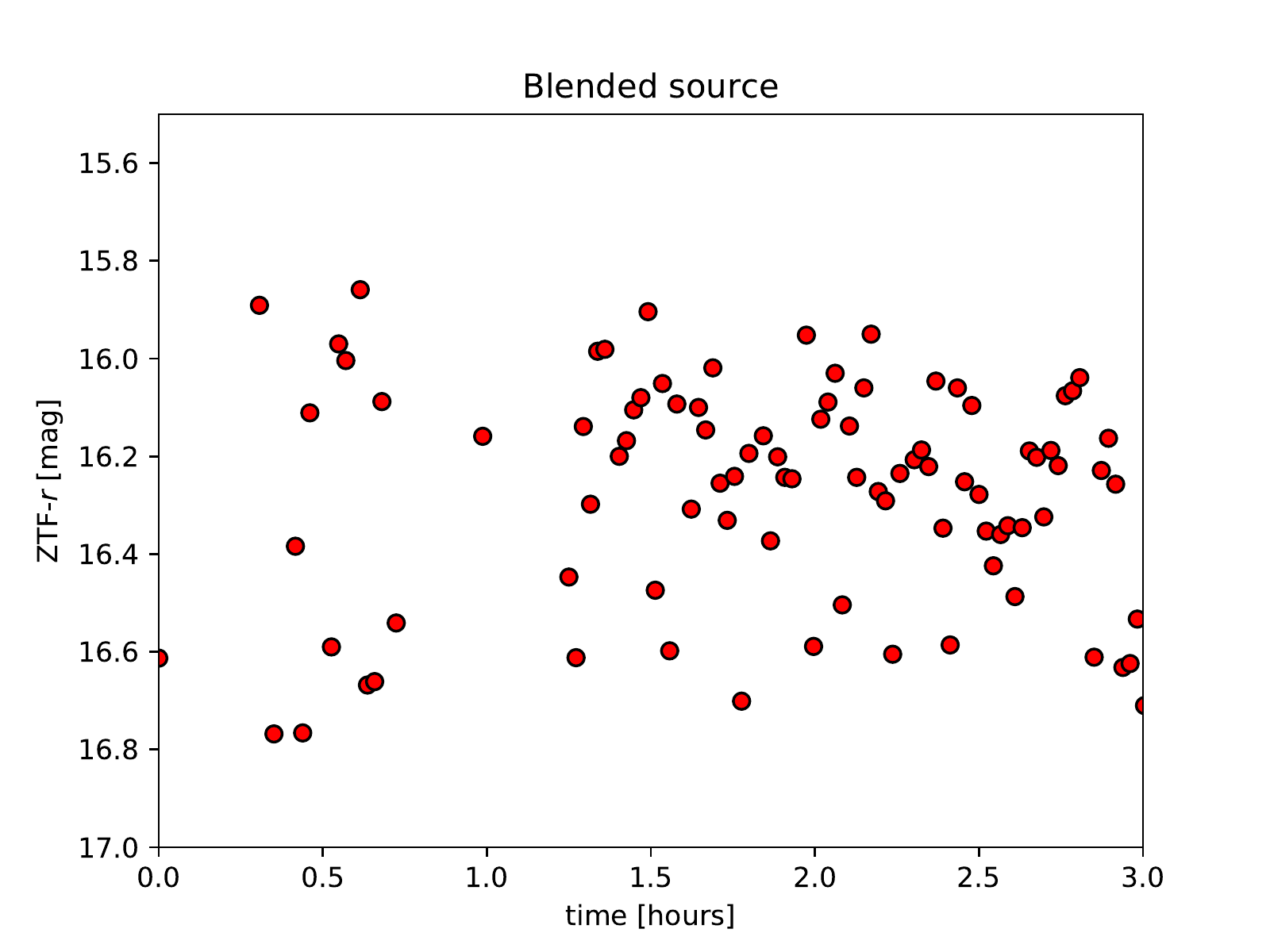} 
\includegraphics[width=0.49\textwidth]{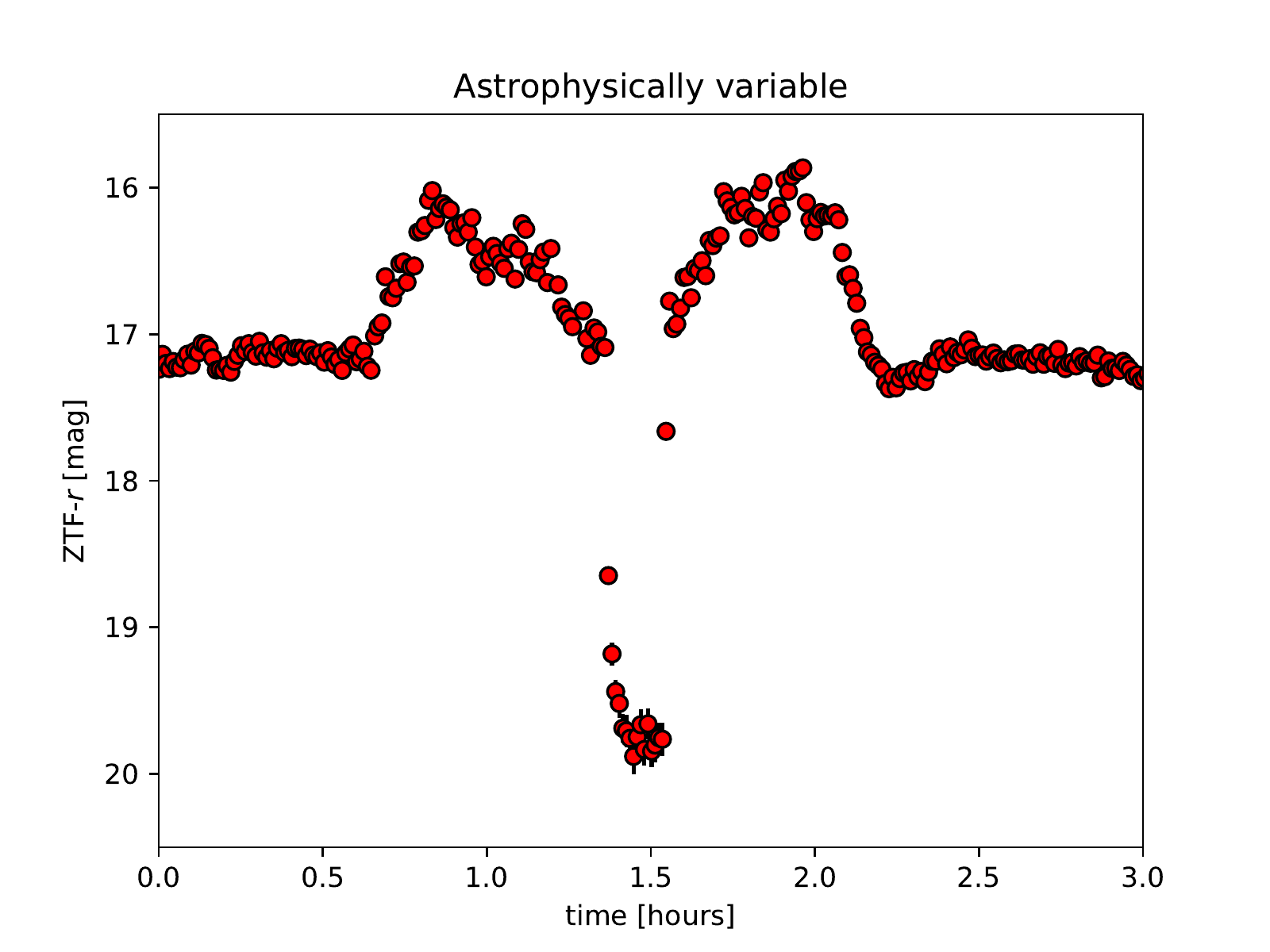}
\includegraphics[width=0.49\textwidth]{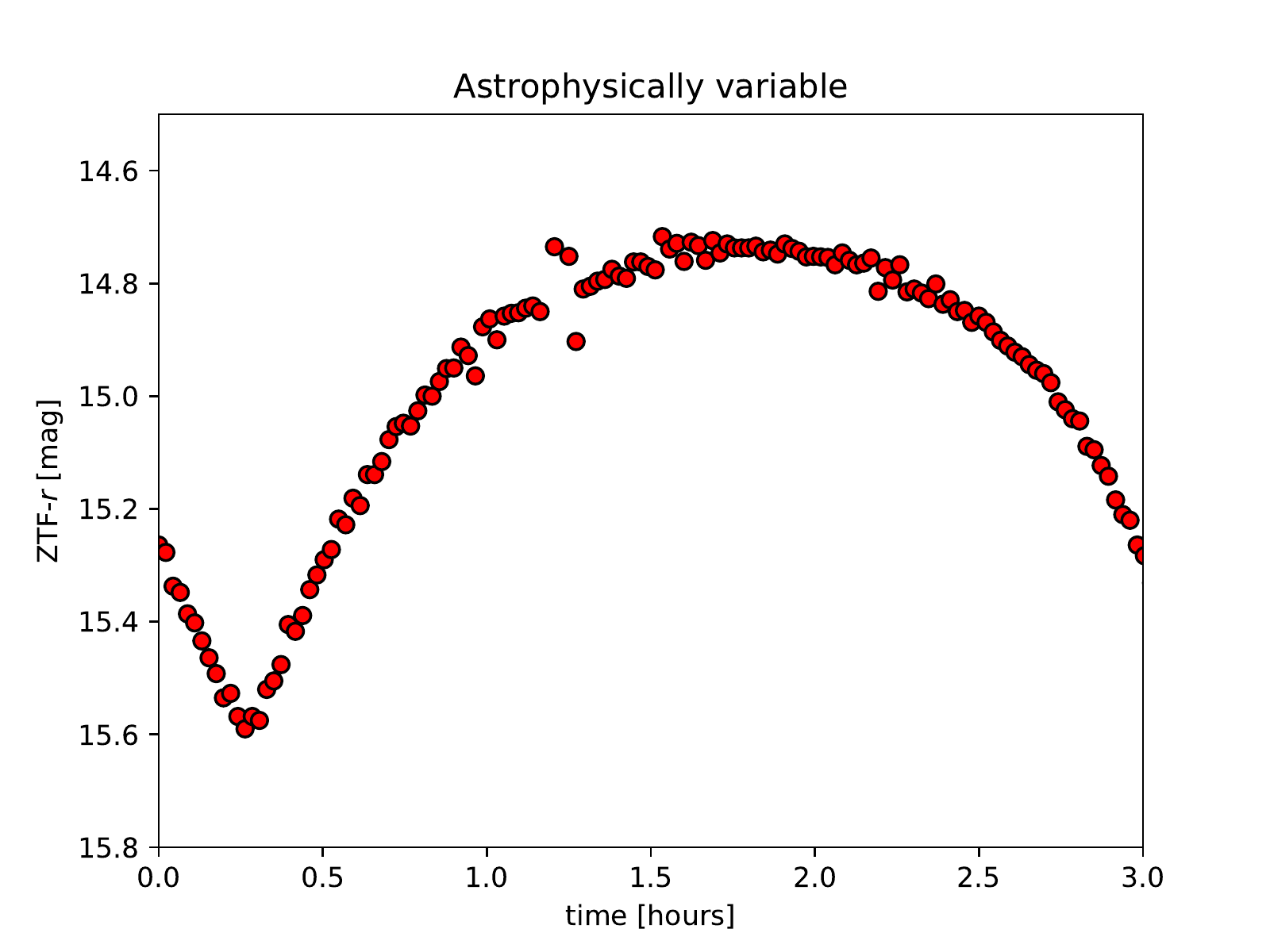}
    \caption{Example light curves of blended sources (upper panels) and real astrophysical variable sources (lower panels).}
   \label{fig:ztf_blend}
\end{figure*}

Different statistics can be used to evaluate whether an object is considered variable or not. We decided to use the interquartile range (IQR) which corresponds to the difference between the upper and lower quartile value. The main advantage of IQR is that it is a robust statistic not affected by extreme outliers due to individual bad photometry data points and therefore a better estimate of the general spread around the median magnitude. When there are outliers in a sample, the median and IQR are best used to summarize a typical value and the variability in the sample, respectively. Large variability in the light curve leads to a larger IQR.

\begin{figure*}
\centering
\includegraphics[width=0.49\textwidth]{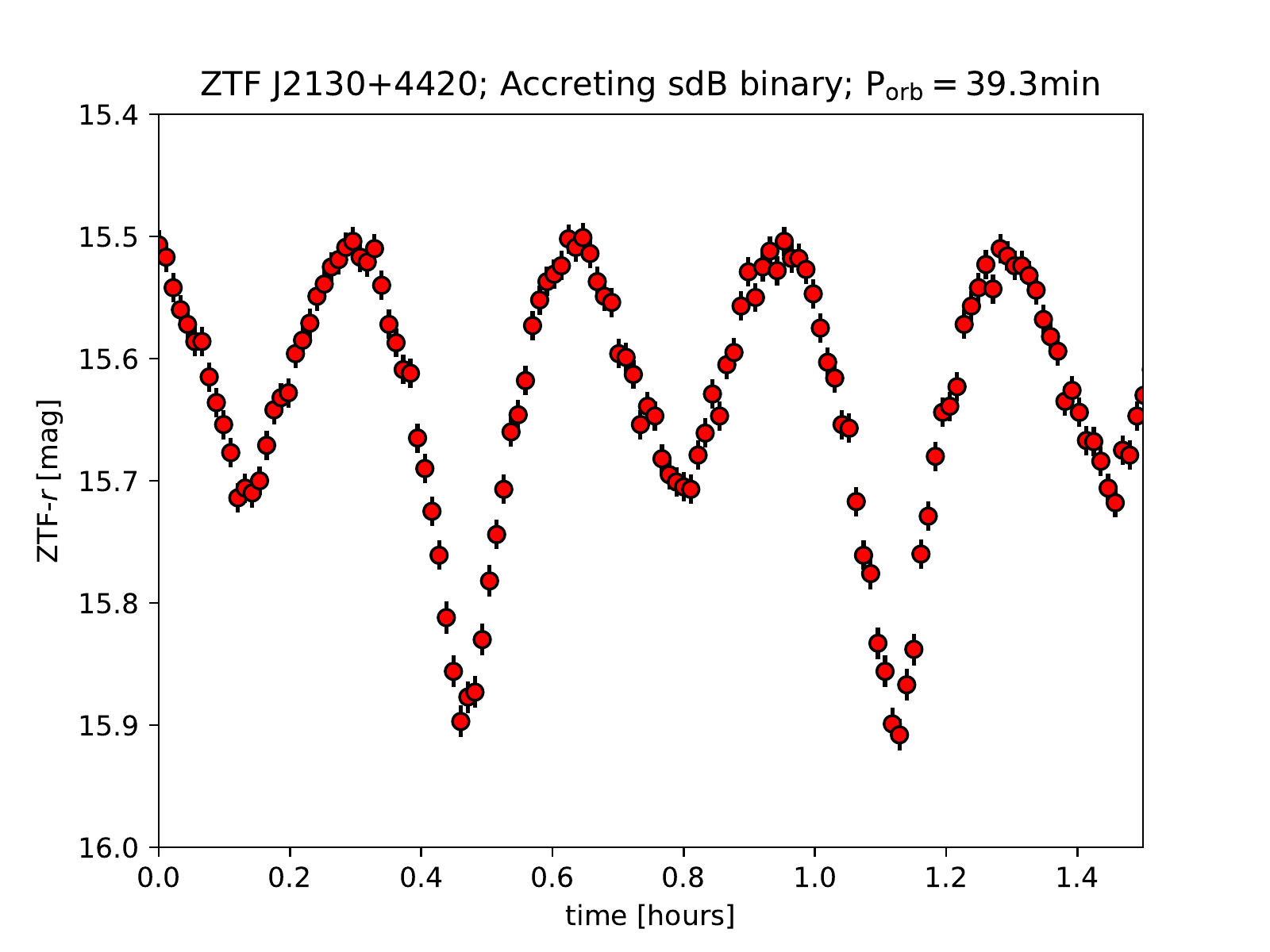} 
\includegraphics[width=0.49\textwidth]{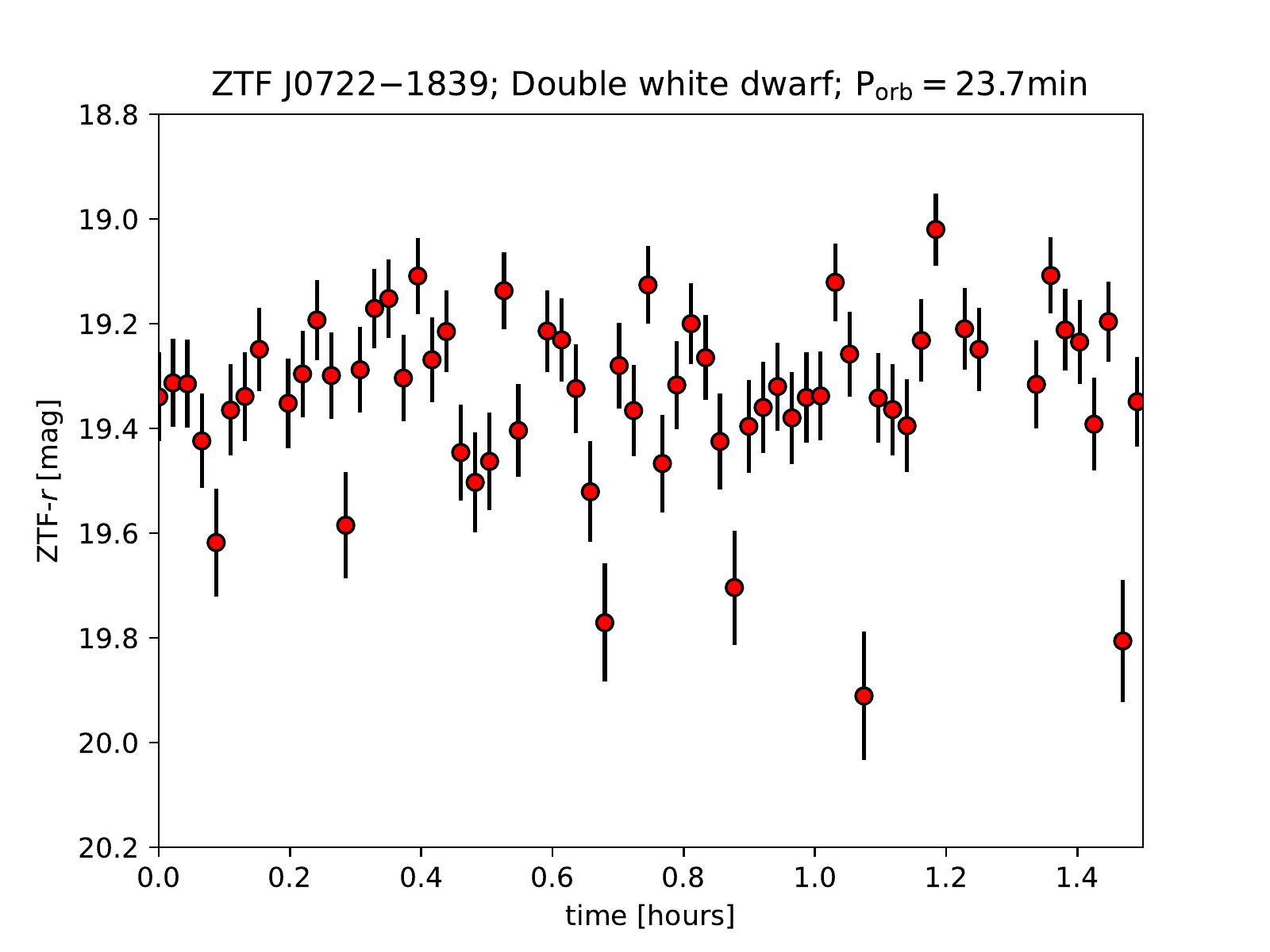} 
\includegraphics[width=0.49\textwidth]{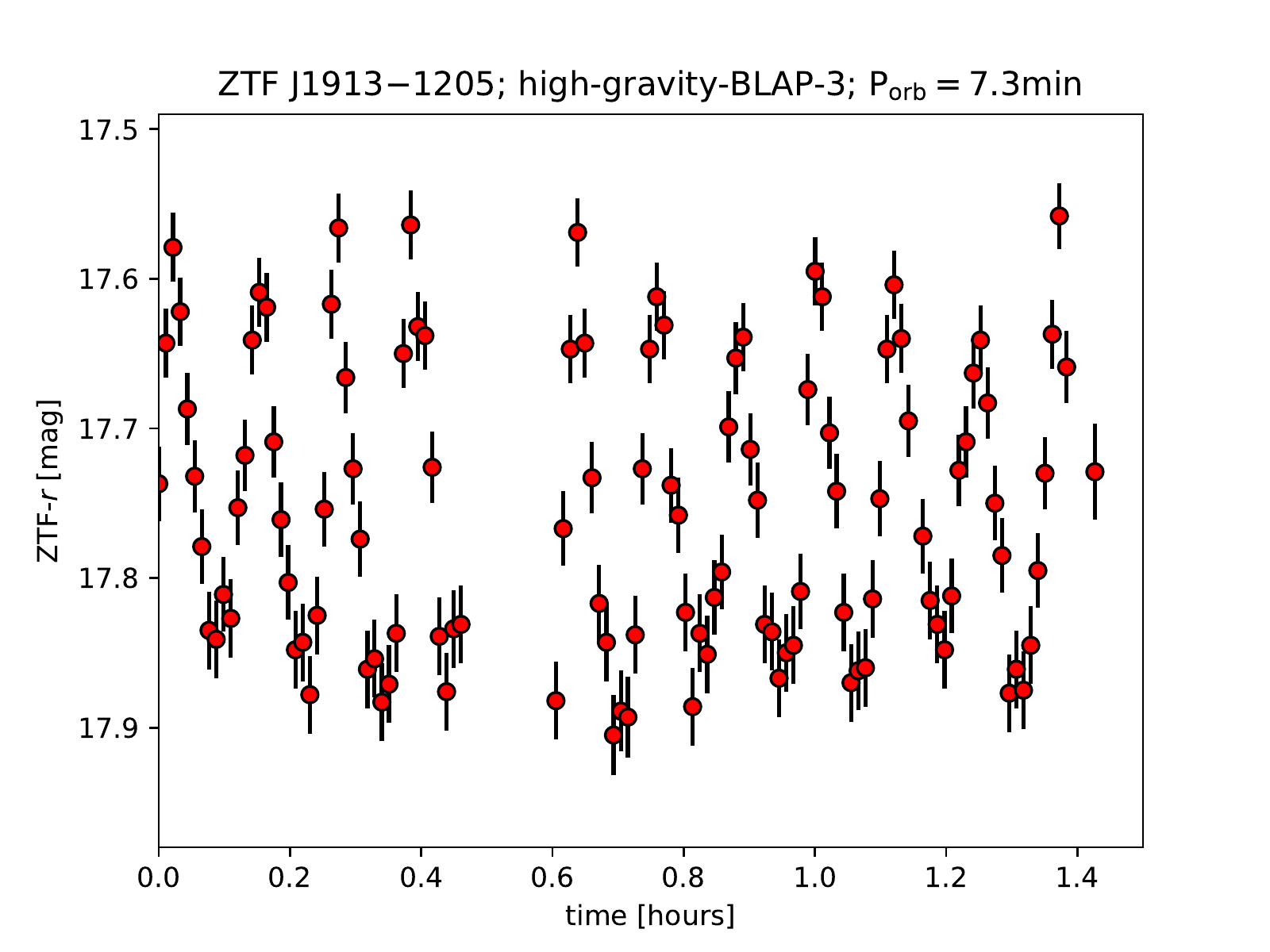}
\includegraphics[width=0.49\textwidth]{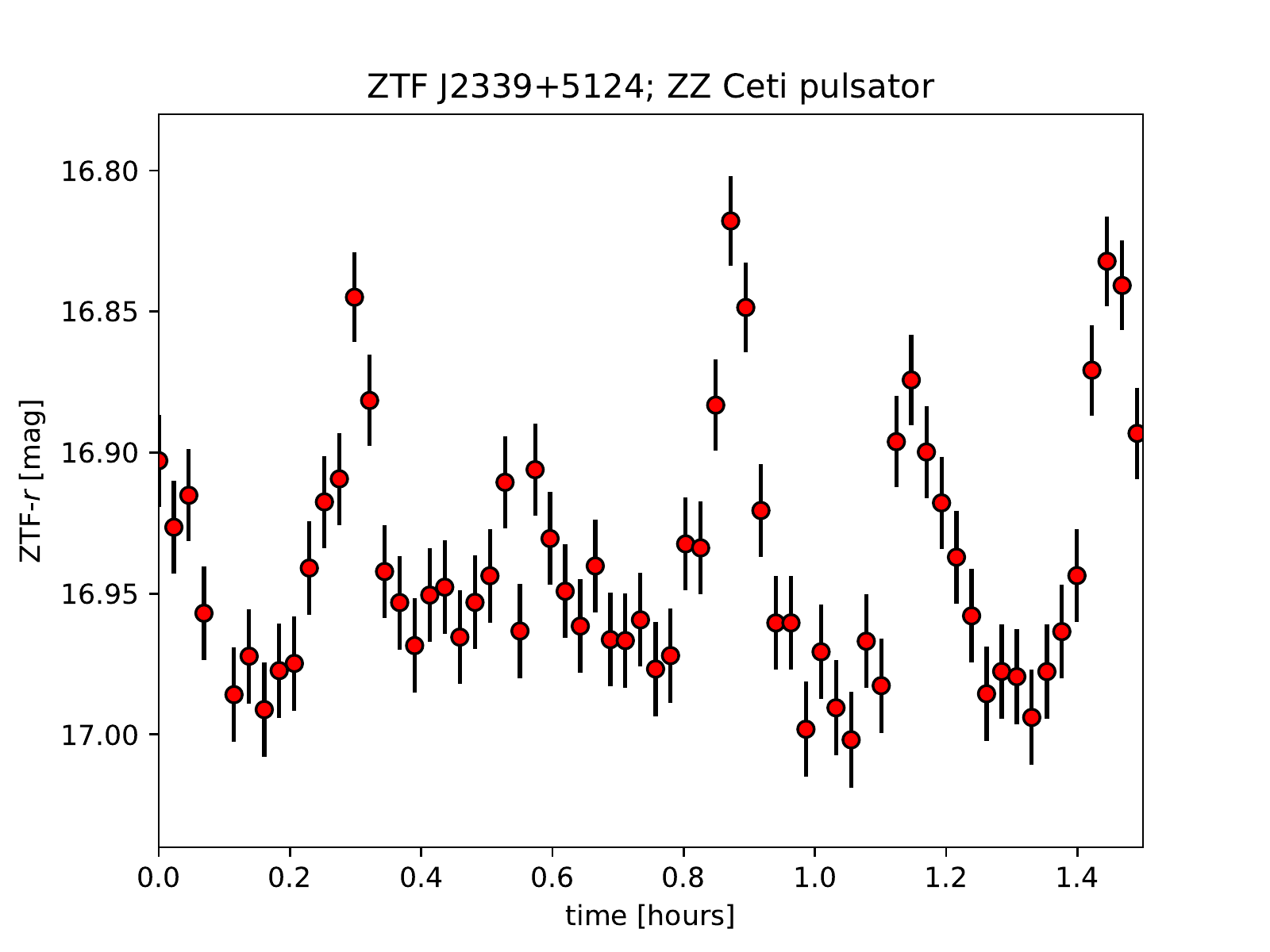}
\includegraphics[width=0.49\textwidth]{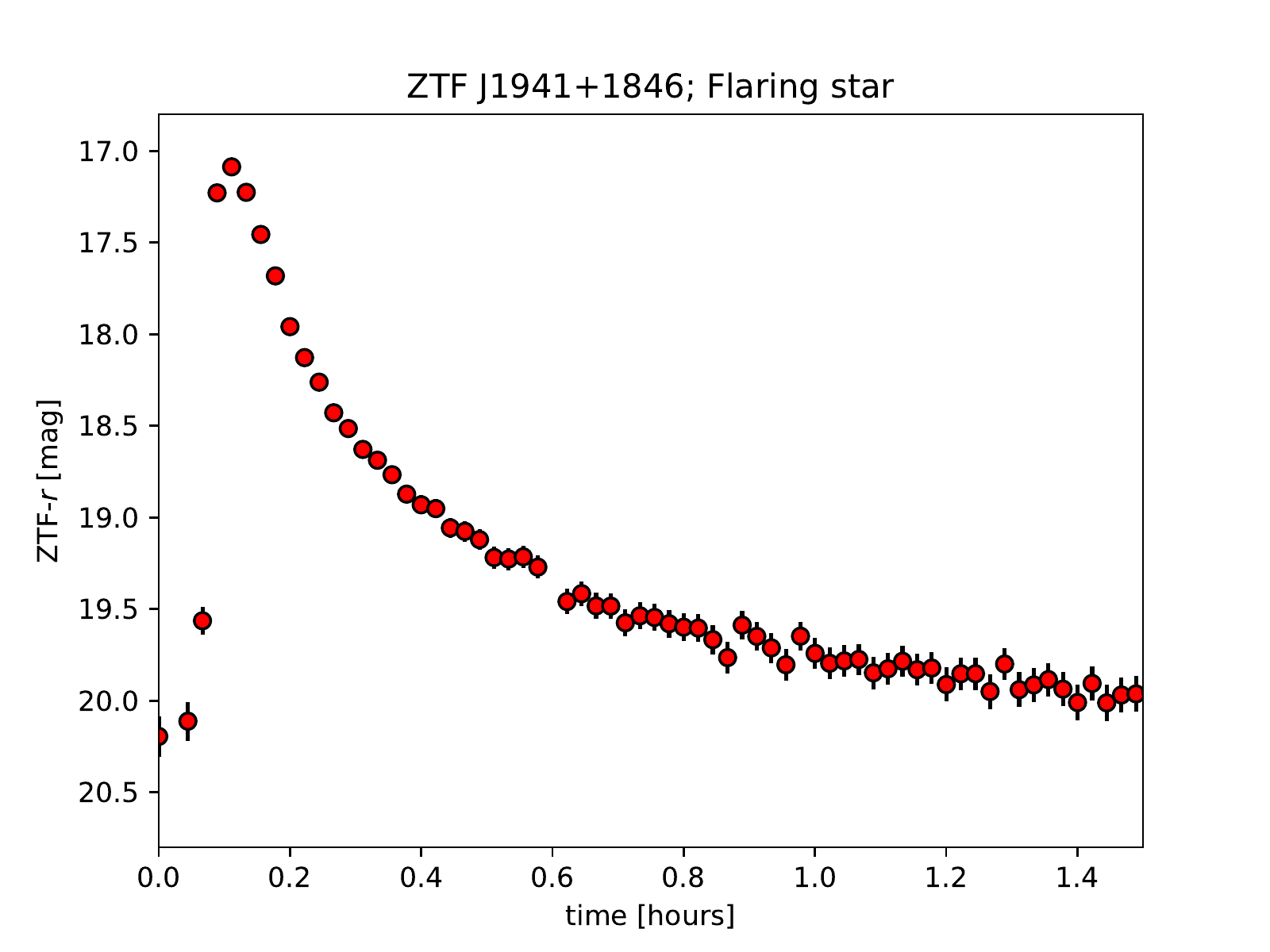} 
    \caption{Example light curves of different types of variable objects discovered in the high-cadence Galactic Plane survey. {\bf Upper left panel:} Accreting sdB binary \citep{kup20}. {\bf Upper right panel:} Eclipsing double white dwarf \citep{bur20}. {\bf Middle left panel:} Compact radial mode pulsator \citep{kup19}.  {\bf Middle right panel:} ZZ Ceti pulsator \citep{gui20}. {\bf Lowest panel:} Flaring star.}
   \label{fig:ztf_example}
\end{figure*}

To estimate whether an object is variable in the high-cadence Galactic Plane data we used the following approach. We calculate the median ZTF-$r$ band magnitude for each object. The full magnitude range of all objects in a ZTF field is binned into several bins and each object is added to a bin based on its median magnitude. We used bin sizes of 0.5\,mag between 12 and 16.5\,mag where the IQR is constant at $\approx0.02$\,mag for non-variable sources. Above 16.5 mag we used bin sizes of 0.2\,mag because the IQR for non-variable sources is increasing with increasing magnitudes to $\approx0.2$\,mag at the faint end at ZTF-$r$=20.5\,mag because the underlying noise of the sources is increasing for fainter sources. Each bin contains at least a few thousand objects.

For each magnitude bin, we then calculate a histogram of IQR values. If there are only constant sources with a similar noise pattern, a normal distribution in an IQR histogram is expected. An excess of variable light curves will result in a deviation from a normal distribution towards more objects with larger IQR values. To estimate the excess and number of variable objects we fit a Gaussian to each IQR histogram and define an object as variable if, in a given histogram bin, less than $5\,\%$ of the sources in that magnitude bin are considered constant based on the Gaussian fit. Figure\,\ref{fig:ztf_histo} shows two examples of IQR histograms with Gaussian fits of FieldID=309 and FieldID=331 for the magnitude range 15.5 - 16\,mag. The orange curve corresponds to the Gaussian fit of the underlying histogram and the red dashed line corresponds to the IQR limit above which we call an object a variable.

Using this method we find $15\,\%$ variable sources for FieldID=331, $9\,\%$ variable sources for FieldID=769 and FieldID=538 and $5\,\%$ variable sources for FieldID=309 (see Fig.\,\ref{fig:ztf_variab}). The number of variable sources per field decreases with the stellar density of the field and it is likely that blending could produce false positives where the object appears variable due to a close-by source which contaminates the same pixel. To test that we inspected by eye 300 light curves which show the largest IQR for FieldID=331 which has the highest stellar density and FieldID=309 which has the lowest stellar densities of the selected fields. We find that astrophysically variable sources show a trend in the light curve, whereas non-astrophysically variable sources show typically rapid exposure-to-exposure variations. These variations can be explained in most cases by blending. Fig.\,\ref{fig:ztf_blend} shows two examples of non-astrophysically variable sources and two examples of astrophysically variable sources. We find that out of the selected light curves only $15\,\%$ for FieldID=331 and $40\,\%$ for FieldID=309 show real astrophysical variability which shows that in the higher density fields the contamination of false positives is significantly larger. This is also consistent with the larger spread in IQR values seen in FieldID=331 compared to FieldID=309 (Fig.\,\ref{fig:ztf_histo}) because a larger number of false positives would lead to a larger spread of IQR values. Combining the variable sources with the fraction of real astrophysical variable sources we find that $\approx1-2$\,\% of all sources show significant astrophysical variability in the high-cadence Galactic Plane data. We extracted $\approx230$\,million light curves and therefore expect a few million sources which show astrophysical photometric variability during the high-cadence Galactic Plane observations.

\begin{table*}[t!]
\centering
	\caption{Photometric properties of the high-gravity-BLAPs}\label{tab:photresults}
  \begin{tabular}{lcccccccc}
\hline
		 Object & RA (J2000) & Dec (J2000) & $P$ &  $A_{\textrm {ZTF-r}}^b$  & $\varpi^c$  &  $g^a$ & $g-r^a$  & FieldID  \\ 
		        & ($^{\textrm h}:^{\textrm {min}}:^{\textrm {sec}}$)        & ($^\circ:^\prime:^{\prime\prime}$)  & (sec) & (mmag)     &   [mas]  & (mag) & (mag)  & \\
\hline
ZTF-sdBV1  & 18:26:36.09 &   +10:00:22.3 &  $347.287\pm0.02$ &  $19.9$ & $0.5640\pm0.0372$ & 14.90 & $-$0.41  & 538 \\
ZTF-sdBV2  & 19:13:54.77 & $-$11:05:19.4 &  $360.576\pm0.02$ &  $19.6$ & $0.4568\pm0.0888$ & 16.24 & $-$0.33  & 385 \\
ZTF-sdBV3  & 21:03:08.50 &   +44:43 14.9 &  $365.822\pm0.01$ &  $53.3$ & $0.3550\pm0.0769$ & 16.07 & $-$0.31  & 768 \\
ZTF-sdBV4  & 08:06:07.05 & $-$20:08 19.1 &  $370.481\pm0.02$ &  $26.0$ & $0.4021\pm0.0531$ & 16.37 & $-$0.32  & 311 \\
ZTF-sdBV5  & 21:49:45.56 &   +45:58 36.5 &  $375.690\pm0.02$ &  $36.1$ & $0.3112\pm0.0683$ & 16.55 & $-$0.35  & 770 \\
ZTF-sdBV6  & 07:40:00.71 & $-$14:42 02.5 &  $375.782\pm0.02$ &  $21.1$ & $0.5979\pm0.0394$ & 15.10 & $-$0.36  & 310 \\
ZTF-sdBV7  & 19:24:08.62 & $-$12:58 30.9 &  $377.468\pm0.02$ &  $19.3$ & $0.7399\pm0.0355$ & 14.64 & $-$0.41  & 385 \\
ZTF-sdBV8  & 06:54:48.55 & $-$25:22 08.3 &  $435.459\pm0.02$ &  $31.4$ & $0.5458\pm0.0318$ & 13.93 & $-$0.44  & 259 \\
ZTF-sdBV9  & 18:52:49.07 & $-$18:46 08.4 &  $524.829\pm0.02$ &  $73.4$ & $0.1427\pm0.1538$ & 17.12 & $-$0.38  & 334 \\
ZTF-sdBV10 & 19:49:59.34 &   +08:31 06.1 &  $646.306\pm0.02$ &  $32.8$ & $0.3884\pm0.0642$ & 16.23 & $-$0.38  & 541 \\
ZTF-sdBV11 & 07:18:43.69 & $-$02:29 31.2 &  $688.485\pm0.02$ &  $30.2$ & $0.1571\pm0.1329$ & 17.52 & $-$0.33  & 411 \\
ZTF-sdBV12 & 17:07:41.34 & $-$15:22 42.9 &  $995.479\pm0.02$ &  $37.3$ & $0.4376\pm0.0346$ & 14.11 & $-$0.35  & 330 \\
\hline
\hline
\end{tabular}
\begin{flushleft}
$^a$ from the Pan-STARRS release 1 (PS1) survey  \citep{cham16} reddening corrected using \citet{gre19}\\
$^b$ amplitude in ZTF-$r$ from the ZTF light curves\\
$^c$ parallaxes from Gaia EDR3 \citep{gai16,gai20}
\end{flushleft}
\label{tab:sdbvs}
\end{table*}



\begin{figure*}
\centering
\includegraphics[width=0.325\textwidth]{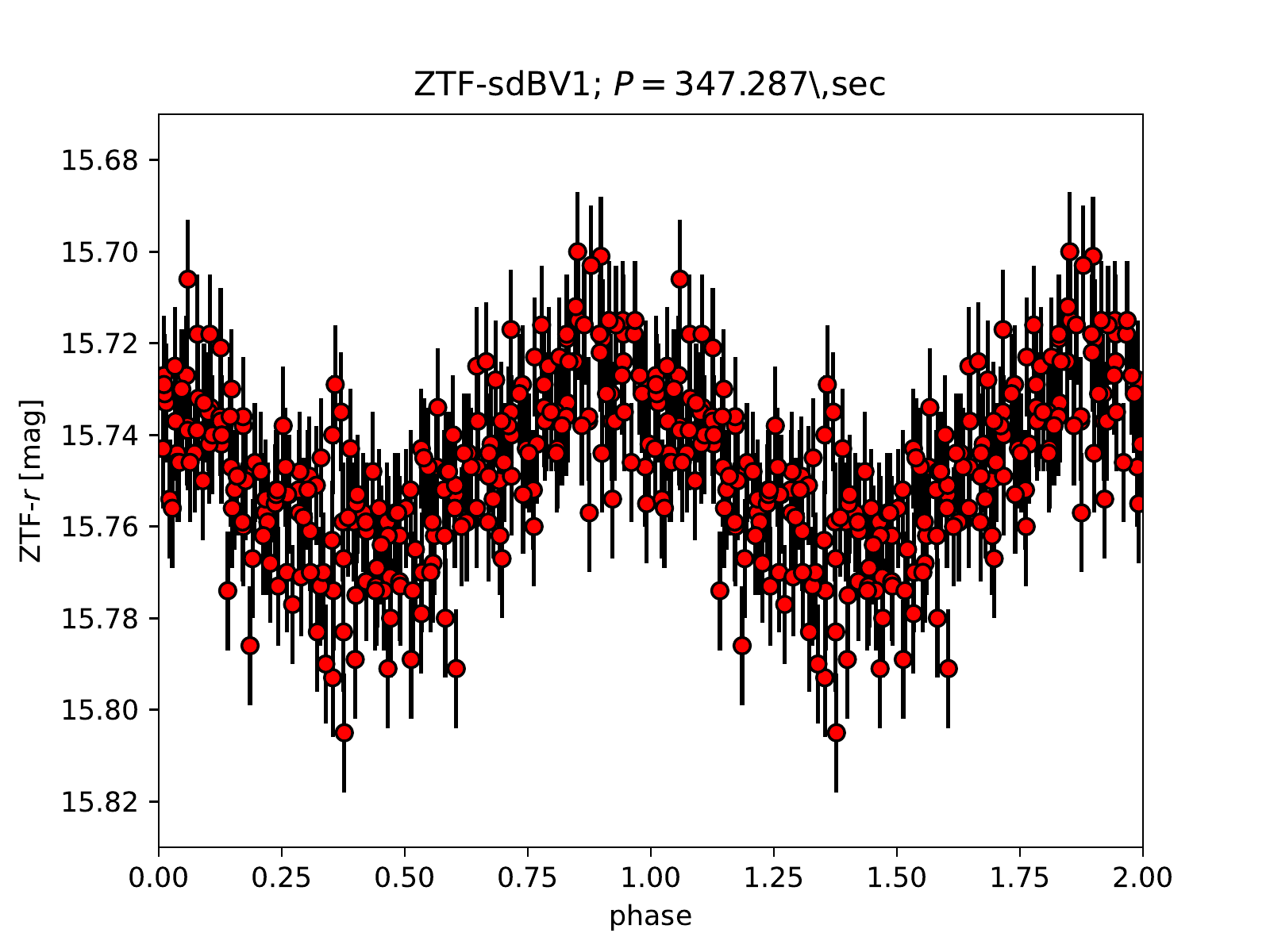} 
\includegraphics[width=0.325\textwidth]{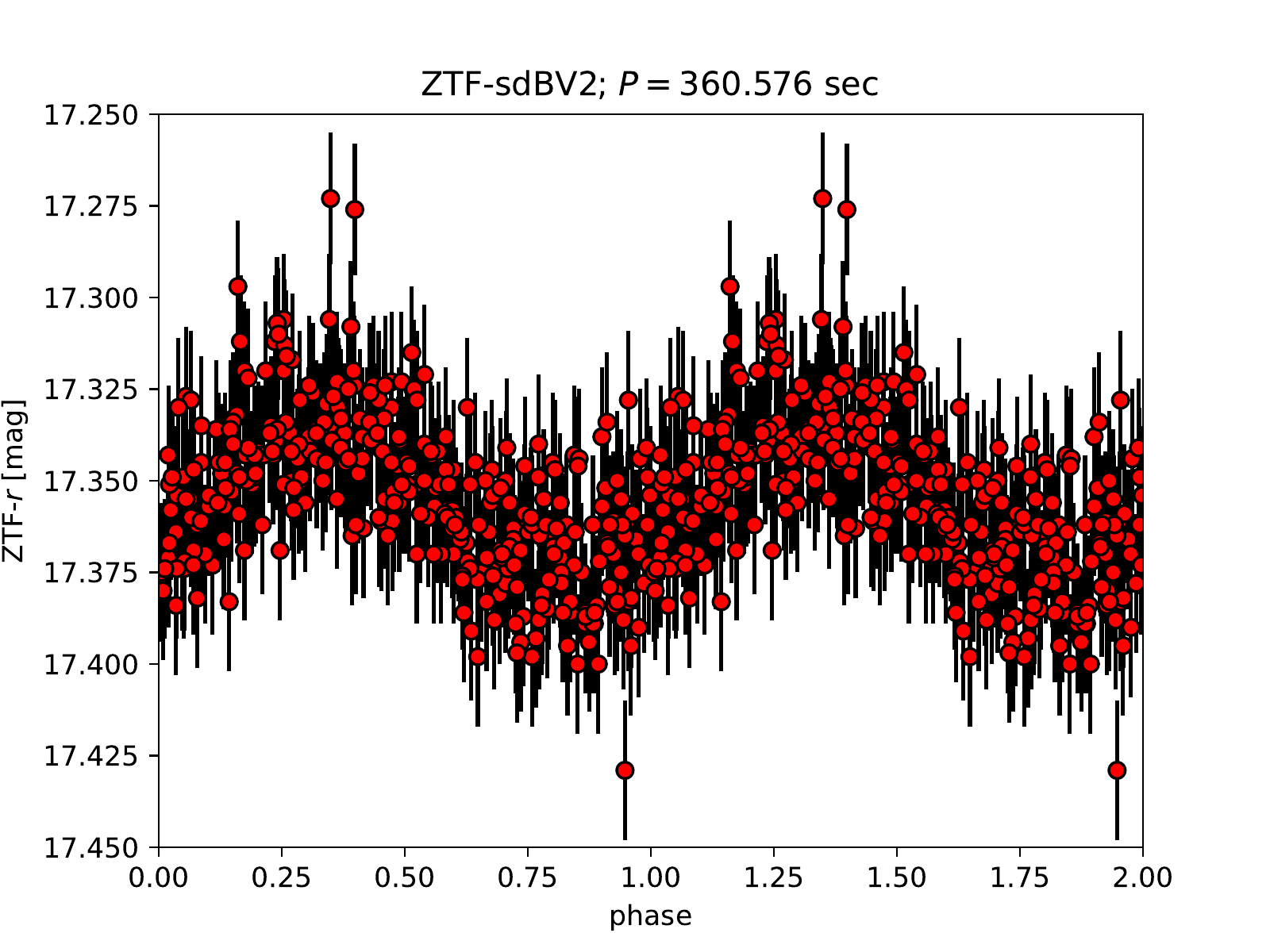} 
\includegraphics[width=0.325\textwidth]{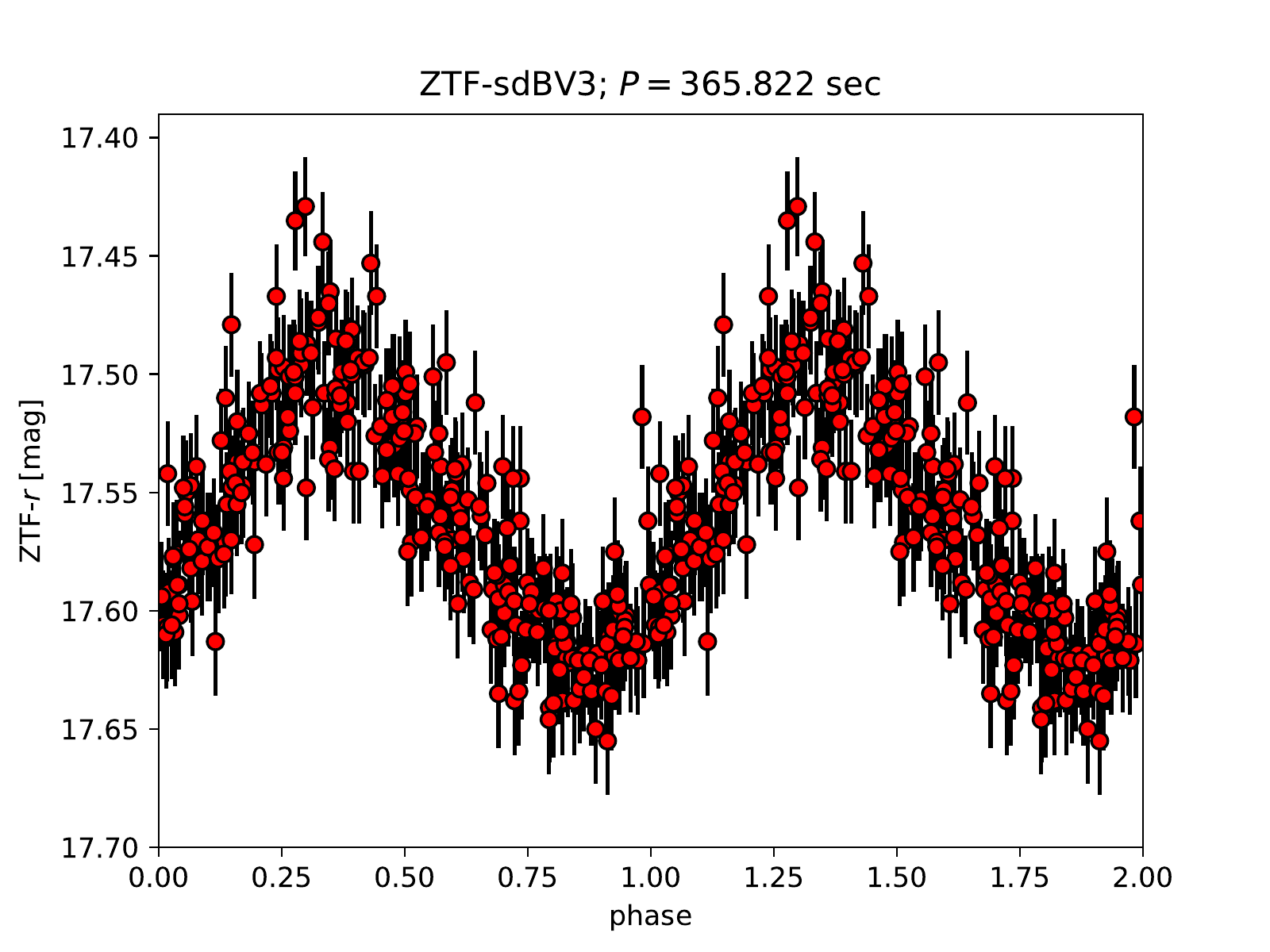}
\includegraphics[width=0.325\textwidth]{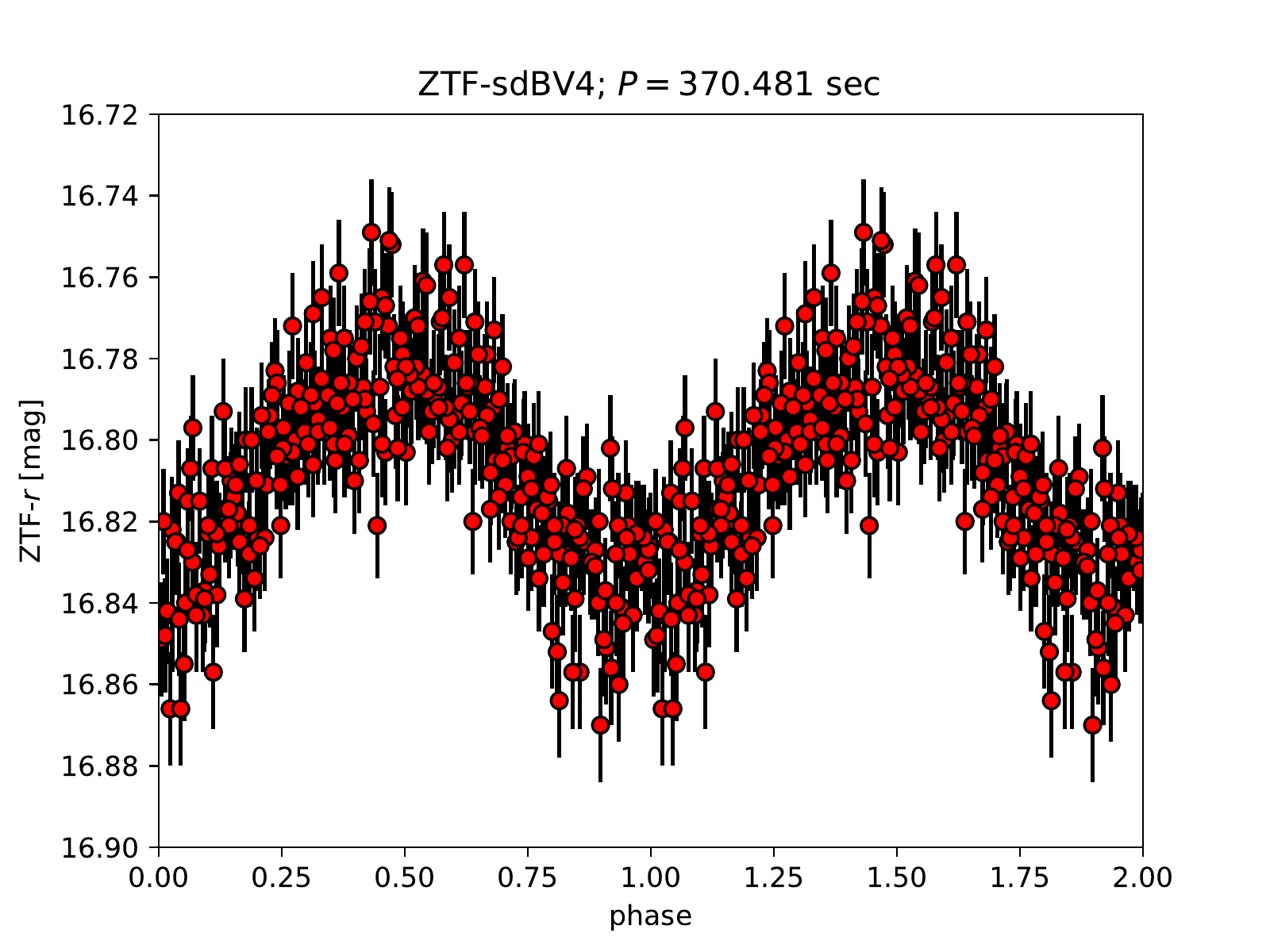} 
\includegraphics[width=0.325\textwidth]{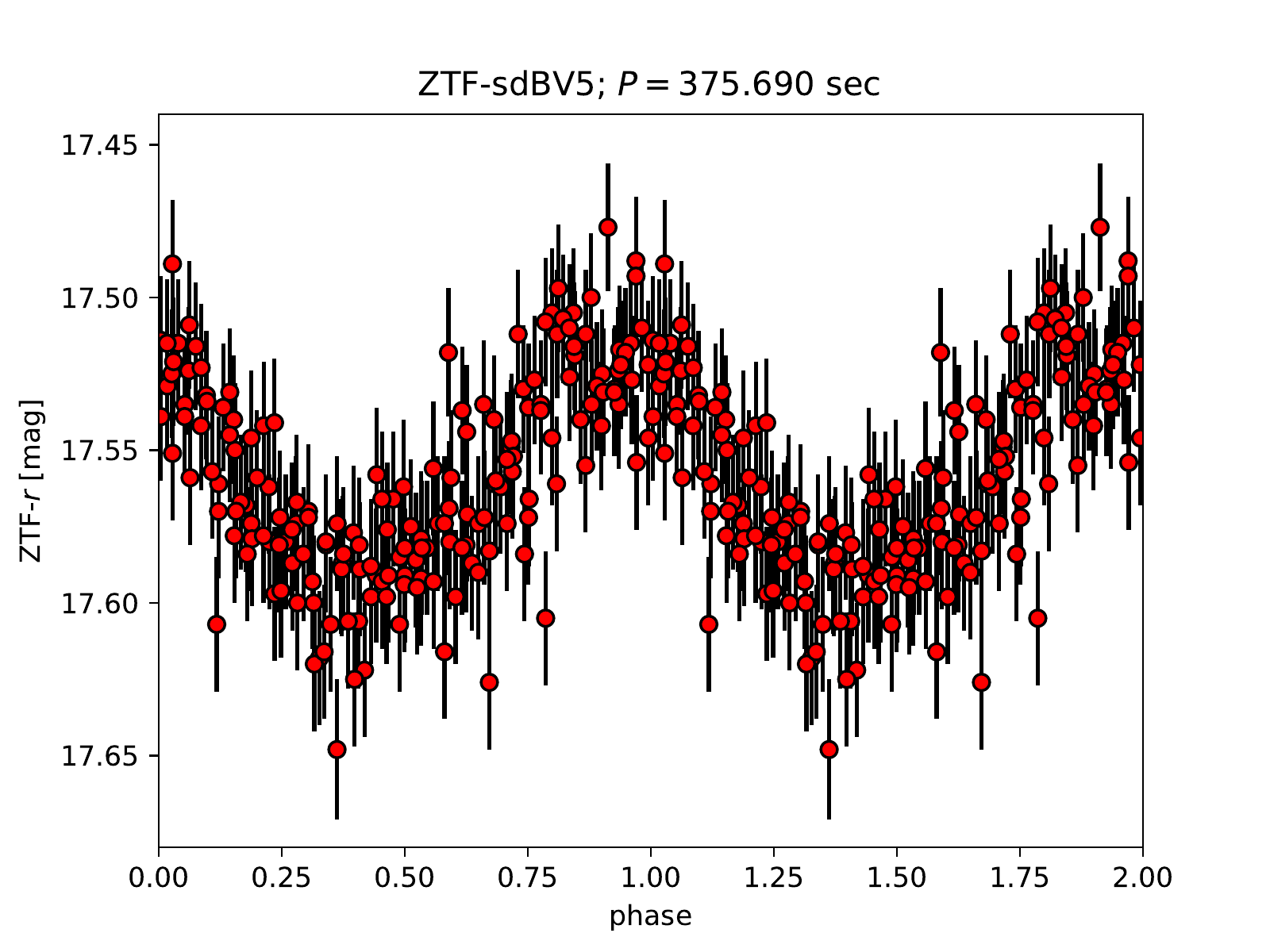} 
\includegraphics[width=0.325\textwidth]{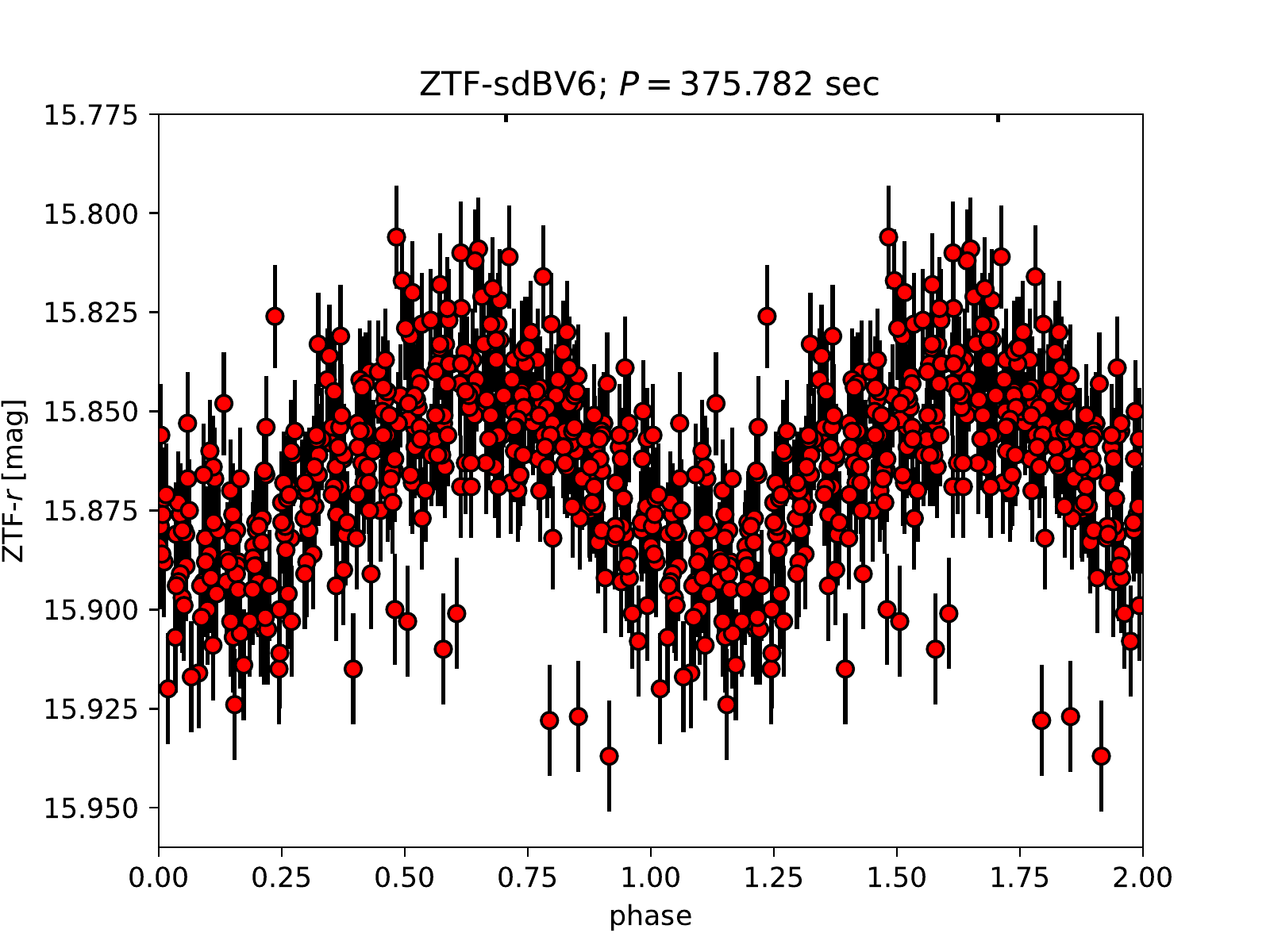}
\includegraphics[width=0.325\textwidth]{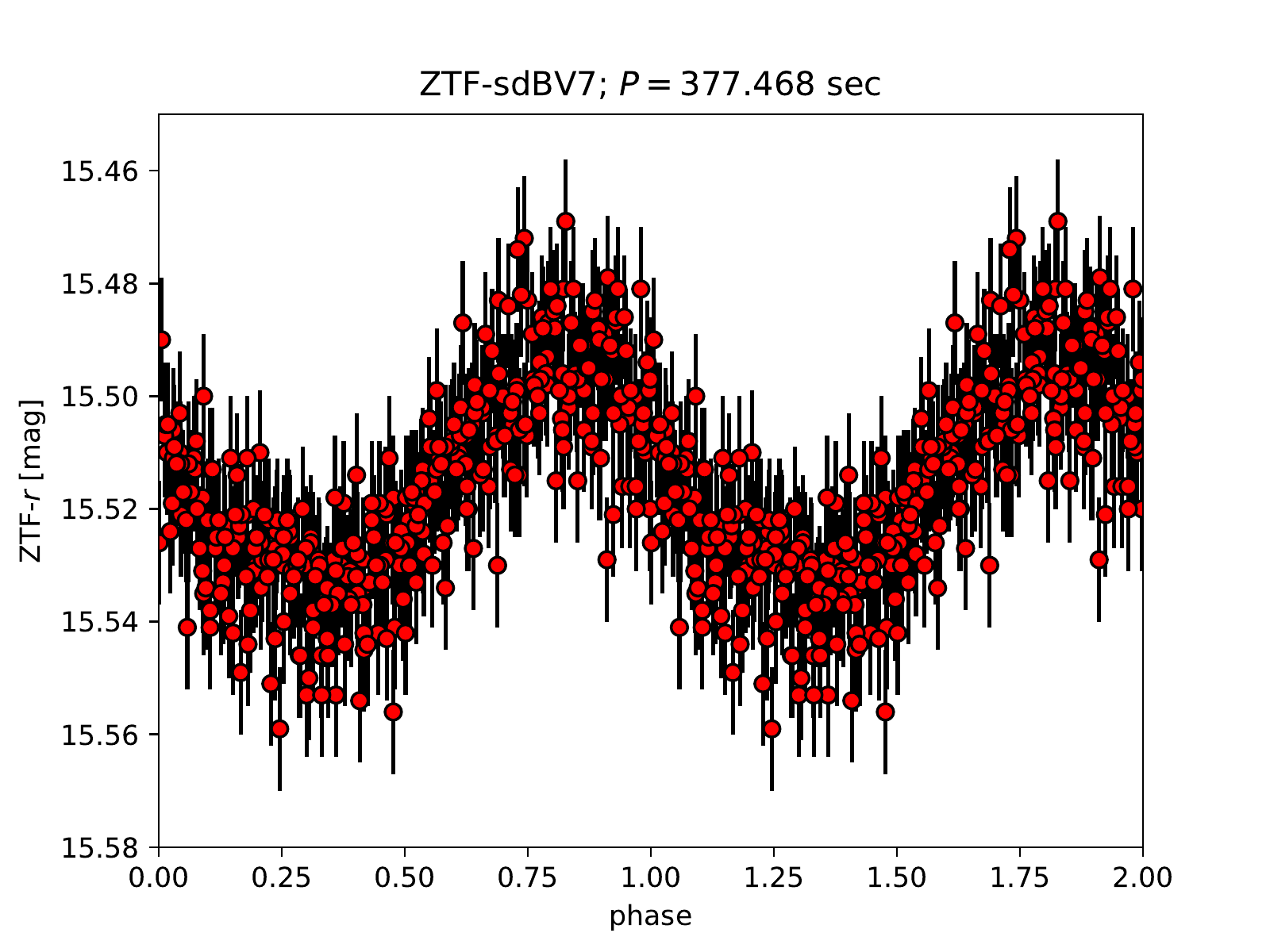} 
\includegraphics[width=0.325\textwidth]{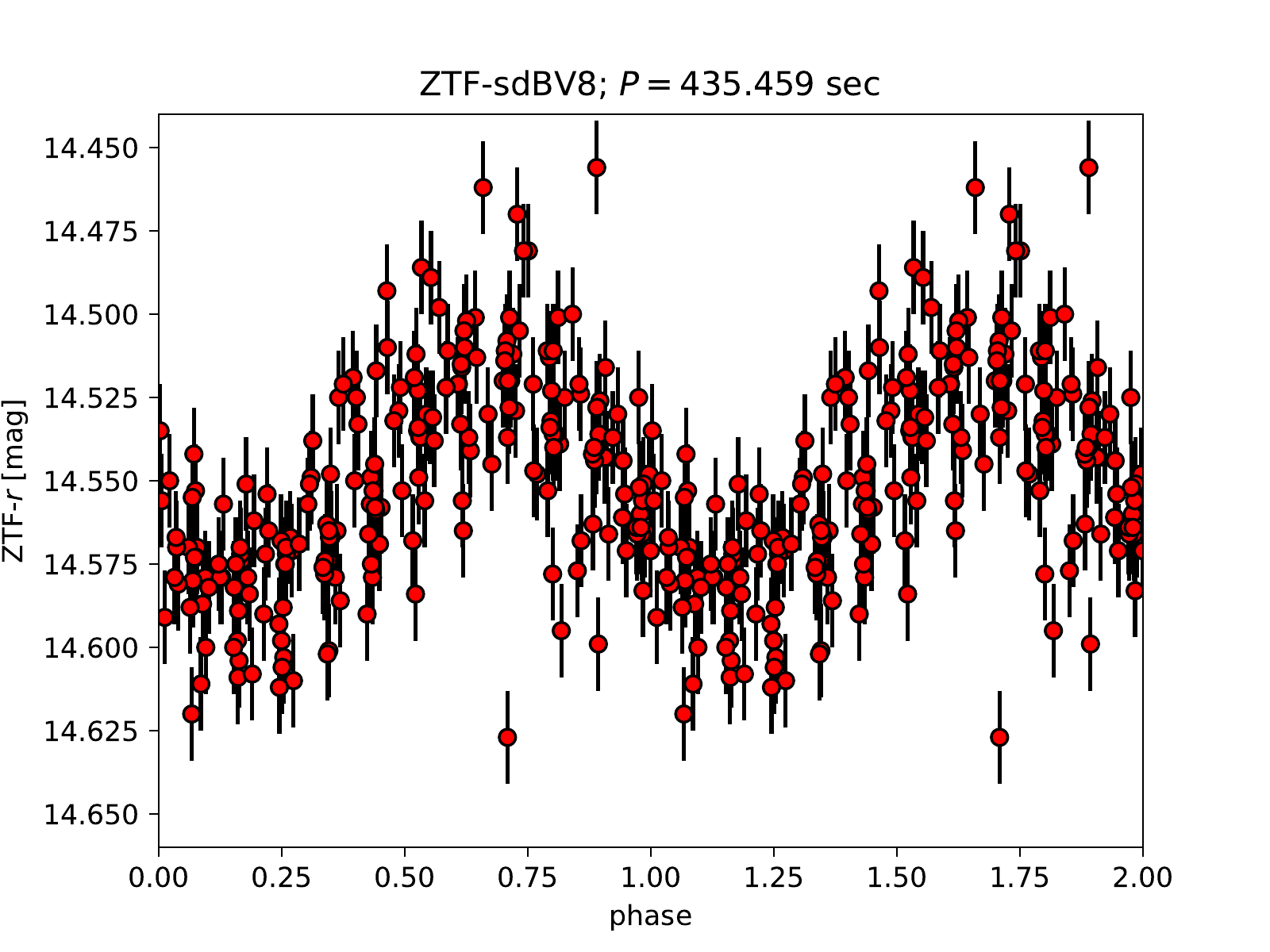} 
\includegraphics[width=0.325\textwidth]{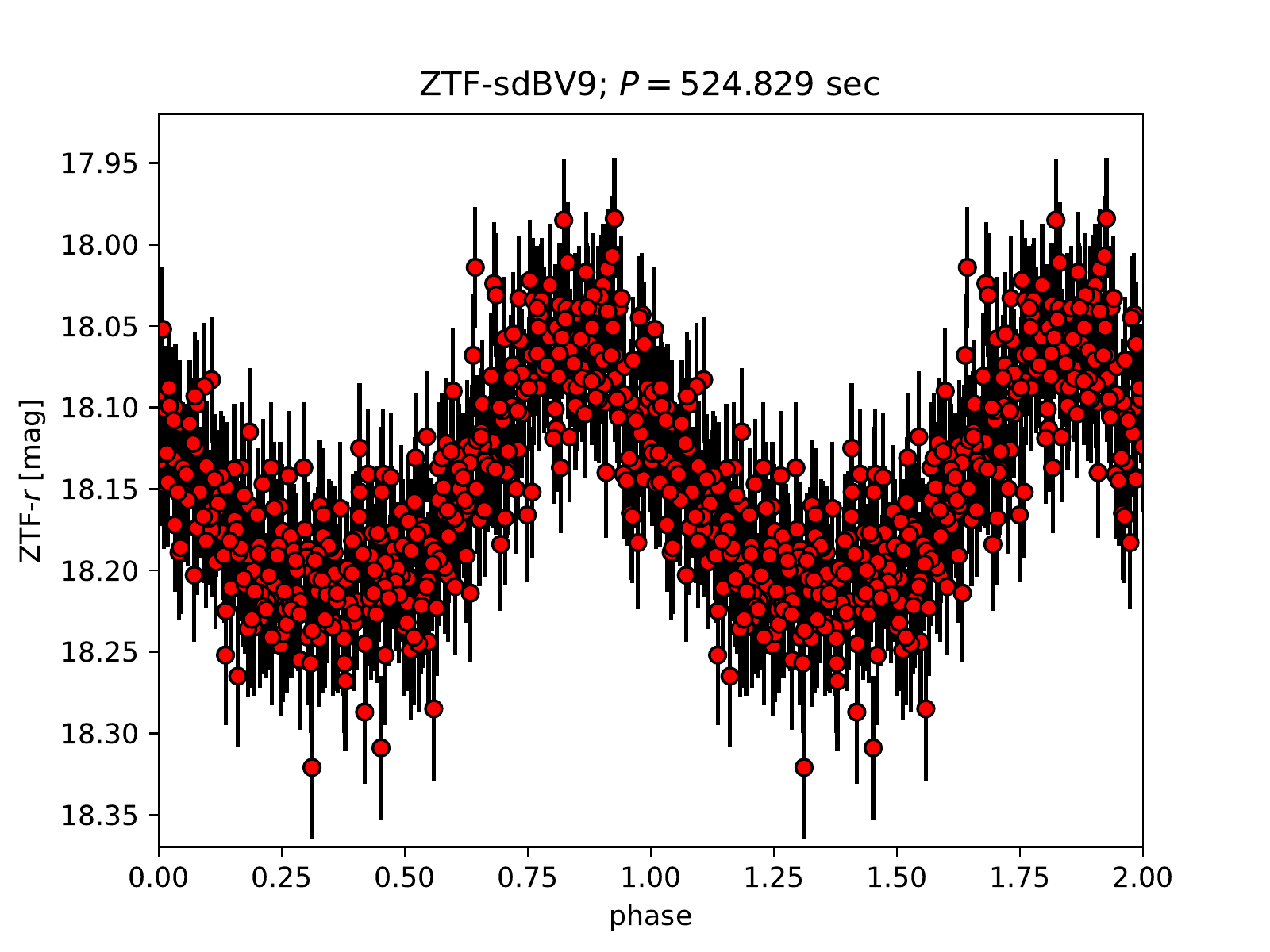}
\includegraphics[width=0.325\textwidth]{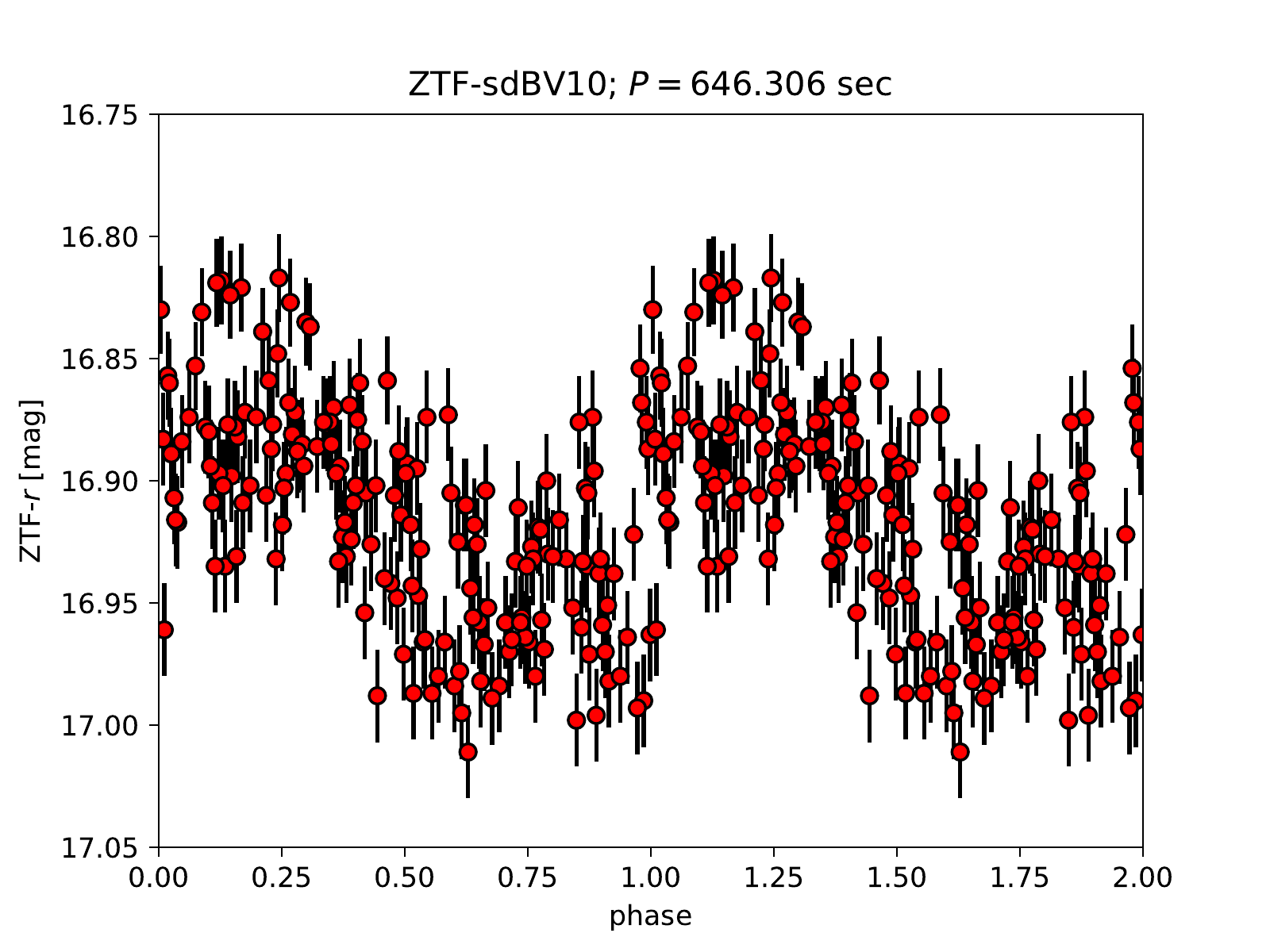} 
\includegraphics[width=0.325\textwidth]{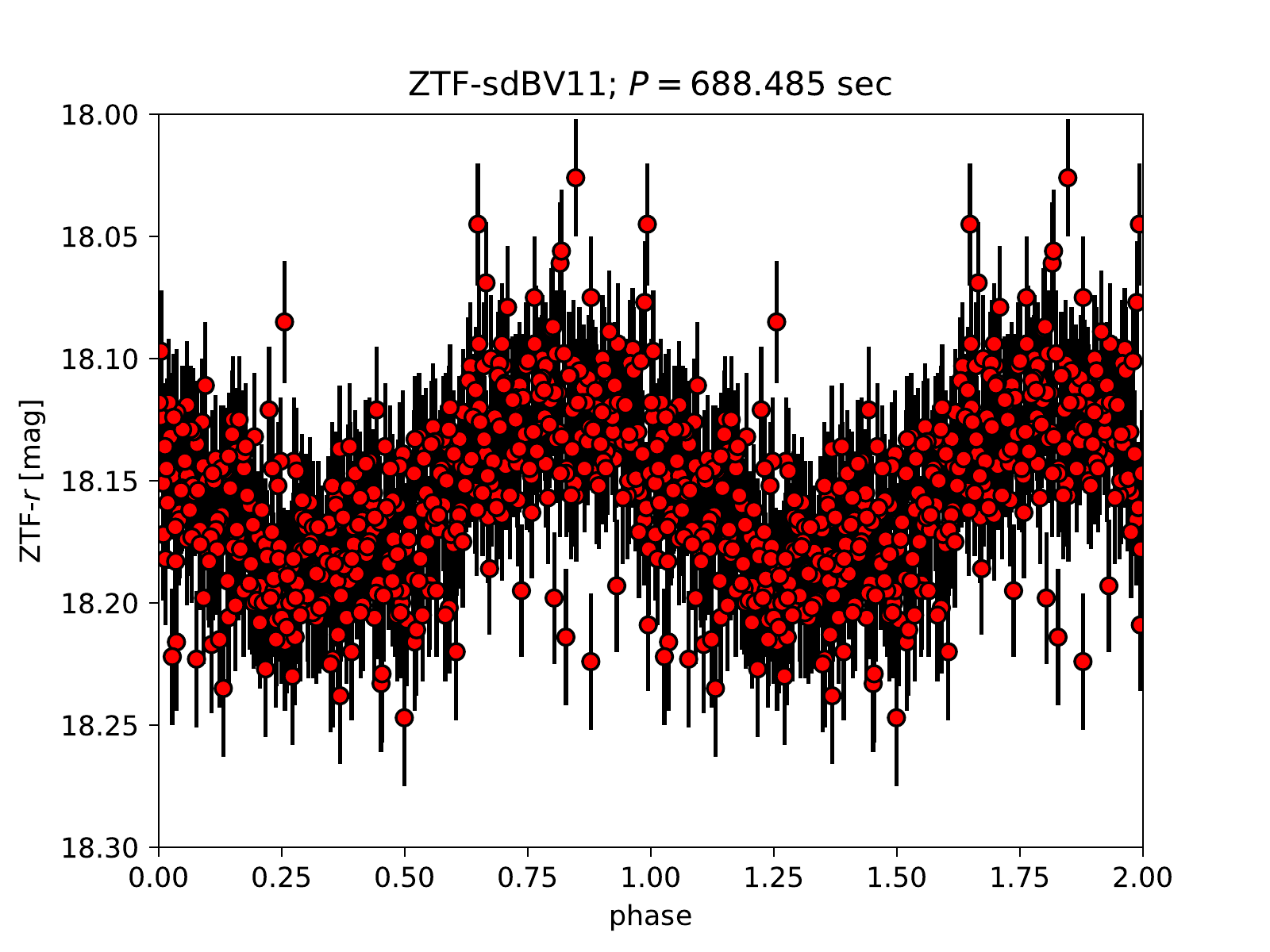} 
\includegraphics[width=0.325\textwidth]{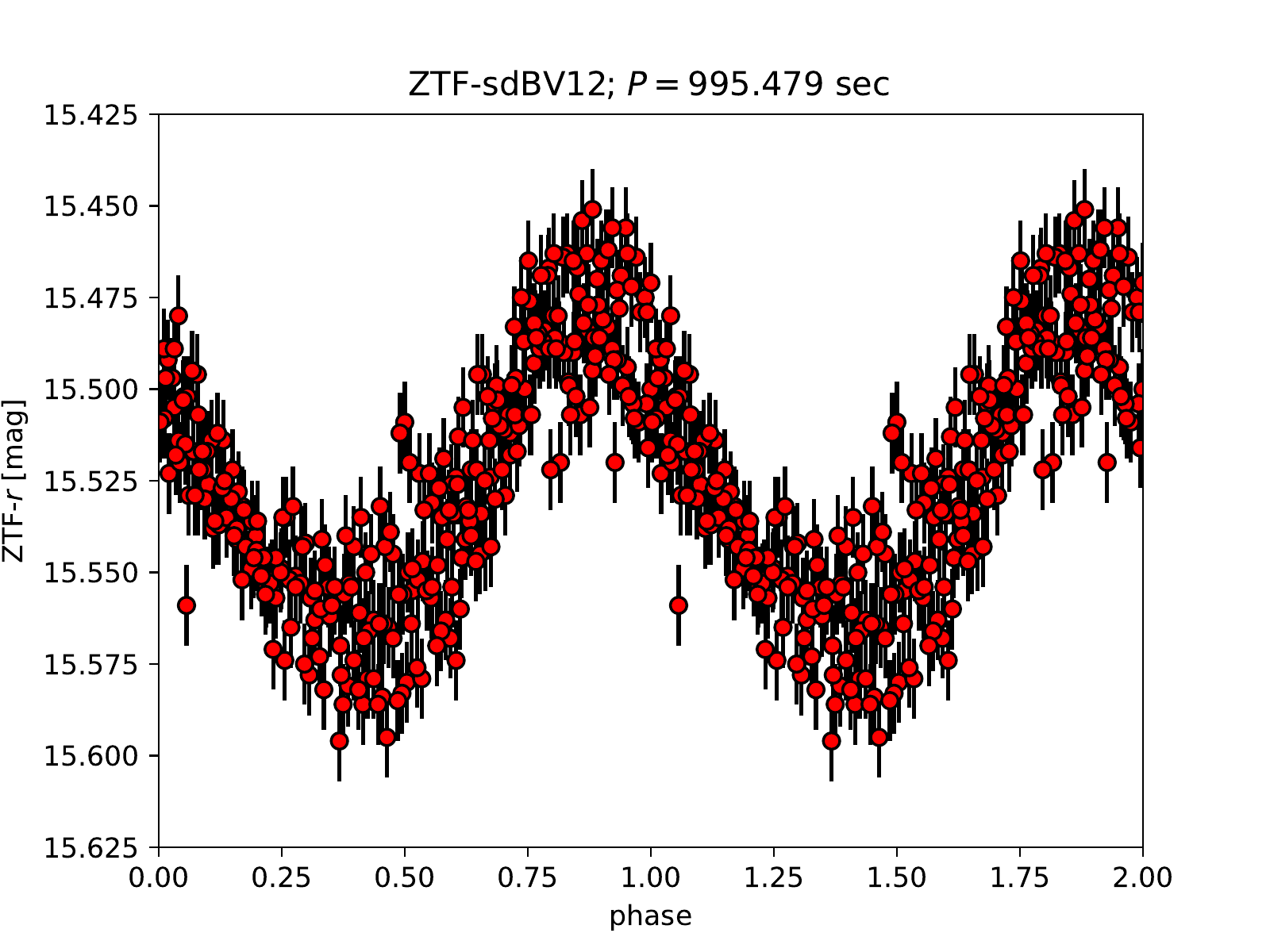}
    \caption{Phase folded ZTF light curves of the newly discovered radial mode sdB pulsators. All systems were discovered as part of the ZTF high-cadence Galactic Plane survey.}
   \label{fig:sdbvs}
\end{figure*}

\section{Progress report}\label{sec:progress}
The primary science driver for the survey is the detection and study of ultracompact binaries and rapid pulsators. Because the data volume is too large to inspect every single light curve we applied different cuts to pre-select light curves in early analysis. Depending on the science goal different selection criteria are required including a color cut using PanSTARRS DR2 colors or a selection based on the position in a color-magnitude diagram using Gaia DR2 \citep{gai16,gai18}. In a targeted search for hot subdwarf (sdB/sdO) binaries we cross-matched the catalogue of 40,000 hot subdwarf candidates \citep{gei19} with the ZTF high-cadence Galactic plane data. These searches have already revealed some interesting discoveries, shown in Fig.\,\ref{fig:ztf_example}.

As part of our cross-match with the hot subdwarf catalog we discovered ZTF\,J2130, the most compact sdB+WD binary and the first member of systems where the sdB fills its Roche Lobe and has started to transfer mass to the WD companion (Upper left panel Fig.\,\ref{fig:ztf_example}; $P_{\rm orb}=39$ min; \citealt{kup20}). In a follow-up search we discovered ZTF\,J2055, a second member of that class \citep{kup20a}. Both objects are only a few hundred million years old and members of the Galactic thin disc population, which explains why these objects have not been seen before and shows the value of a high-cadence Galactic Plane survey.

\begin{figure}
\centering
\centering
\includegraphics[width=0.49\textwidth]{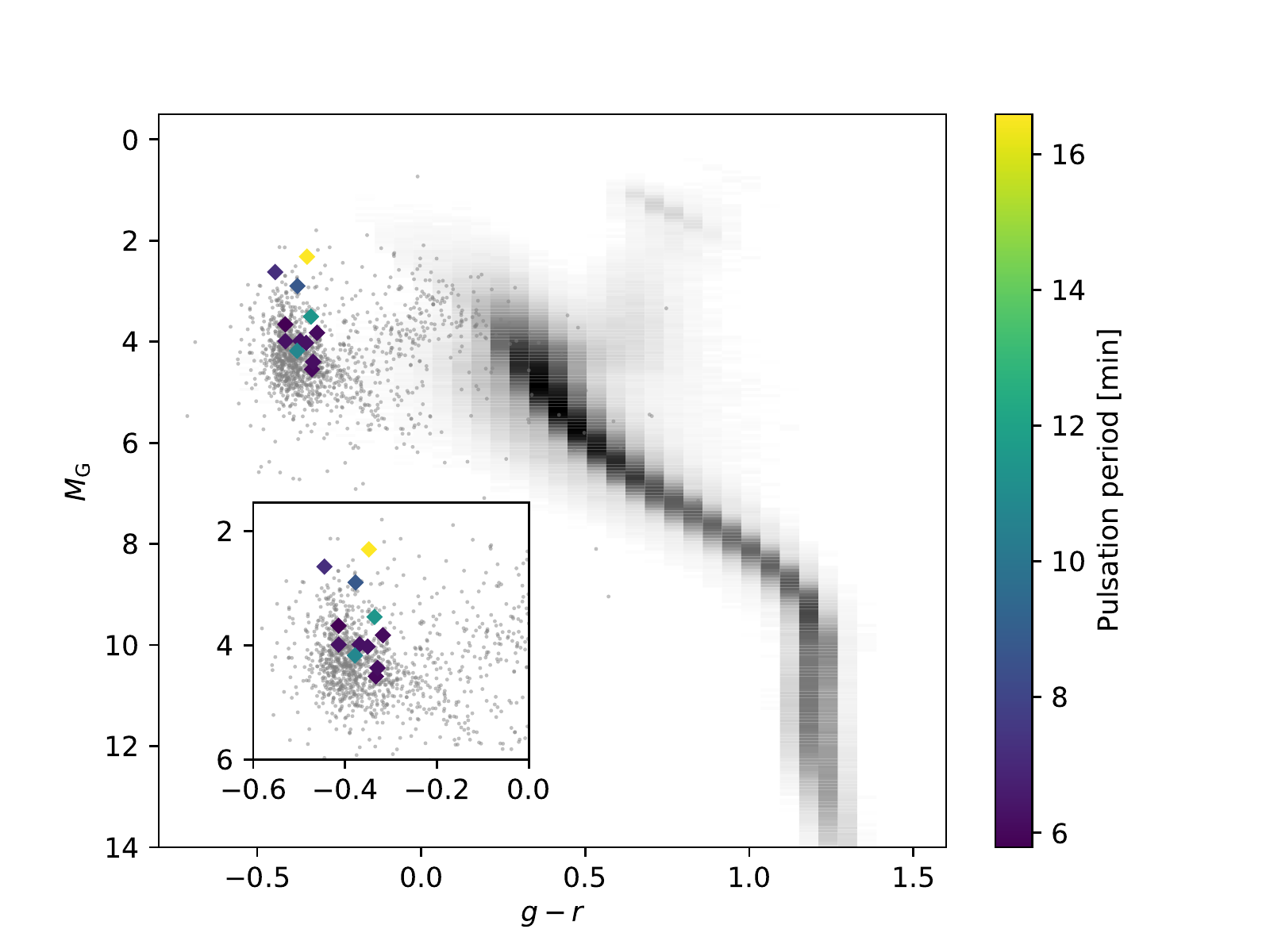} 
    \caption{Hertzsprung Russell diagram of the newly discovered pulsators. The grey dots show known hot subdwarfs selected from \citet{gei20} The grey shaded region corresponds to the underlying Hertzsprung Russell diagram showing the position of the main sequence and the red giant branch.}
   \label{fig:ztf_hr_sdbs}
\end{figure}

\citet{bur20} and \citet{bur20a} presented 12 new ultracompact double white dwarfs including two new AM\,CVn systems, with orbital periods below one hour. Although the high-cadence Galactic Plane survey only covers $\approx$10\,\% of the ZTF sky almost half of the new ultracompact double white dwarfs have been covered by the high-cadence Galactic Plane observations including an 8.8\,min binary \citep{bur20a}. Due to the short eclipse duty cycle of $\leq$10\,\% for some systems, high-cadence Galactic Plane observations are ideal to cover enough in-eclipse points to recover the sources in period finding searches. The upper right panel in Fig.\,\ref{fig:ztf_example} displays an example of an eclipsing double white dwarf binary with an orbital period of 23.7\,min \citep{bur20}. The lower points every 24\,min indicate an eclipse and shows the short duty cycle of some of the eclipsing double white dwarfs.

The survey revealed a new class of large-amplitude radial-mode compact pulsators (three confirmed objects, middle left panel Fig.\,\ref{fig:ztf_example}, \citealt{kup19}). They show typical pulsation amplitudes of $5-20\,\%$ and periods between 3\,min and 8\,min. Although the stars reside in a region in the Hertzsprung Russell (HR) diagram usually occupied by helium core burning sdB stars, their pulsation properties are best explained with low-mass pre-He-WDs currently evolving through this part of the HR diagram contracting towards the He-WD cooling tracks. As such they could be the low mass version of the BLAPs. 

The survey furthermore provided the necessary short-cadence observations to resolve the short-period pulsations of ZZ Cetis. These pulsations are evident in the middle right panel of Figure~\ref{fig:ztf_example}, which highlights ZTF\,J2339+5124, a new ZZ Ceti identified originally in \citet{gui20}. The typical diurnal mean sampling rate of ZTF is too short to resolve the pulsations for many pulsating white dwarfs (ZZ Ceti pulsation modes range from $100 - 1500$\,sec \citealt{muk13,bog20}). However, the Galactic Plane survey's 40\,sec sampling over the span of hours is sufficient to capture consecutive pulsation peaks in ZZ Cetis. These observations may then be coupled with an object's placement relative to the pulsating white dwarf instability strip to allow for its confirmation as a ZZ Ceti (since pulsations are only driven in a narrow range of effective temperature, e.g., \citealt{gia15}), as done for ZTF\,2339+5124.

The survey has also provided a large sample of stellar flare detections. These events are the result of rapid magnetic reconnection for low-mass stars, and typically have timescales of minutes to hours \citep{haw14}, well-matched to the cadence of this survey. We can trace flare events in their entirety (e.g. lower panel Fig.\,\ref{fig:ztf_example}), measuring precise flare energies when the distance is known, as opposed to flares being detected as single-point outliers in the primary ZTF survey.  Using the first data release from the 14 August fields, more than 1500 stellar flare events have been detected (C. Klein et al., in prep). This large sample allows us to study correlations between flare occurrence rates and Galactic height.

\section{Early results - new single-mode hot subdwarf pulsators}\label{sec:early}
Subdwarf B stars (sdBs) are hot stars of spectral type B with luminosities below the main sequence. The formation mechanisms and evolution of sdBs is still debated, although most sdBs are thought to be helium (He)-burning stars with masses $\approx0.5$\,\msol\,and thin hydrogen envelopes (\citealt{heb86,heb09,heb16}). Amongst the sdB stars, two types of multi-periodic pulsators have been discovered, both with generally milli-mag amplitudes. On the hotter side (\teff$\gtrsim28,000$\,K) are the V361\,Hya stars which are pressure mode ($p$-mode) pulsators with typical periods of a few minutes \citep{kil97}. On the cooler side (\teff$\lesssim28,000$\,K) are the V1093\,Her stars which are gravity mode ($g$-mode) pulsators with periods of 45 min to 2 hours \citep{gre03}. Only very few sdBs are know to show amplitudes up to a few percent \citep{ost10}. This includes Balloon 090100001 and PG\,1605+072, which are multi-periodic pulsators with photometric amplitudes of up to 60\,mmag \citep{ore04,til07}. 

Even before their discovery, the variability of sdBs was predicted to be caused by non-radial pulsation modes driven by the opacity bump due to partial ionization of iron \citep{cha96, cha97, fon03}. Until recently the only known single-mode sdB pulsator was CS\,1246, with an amplitude of $\approx30$\,mmag which has decreased below its detection limit over the last few years \citep{bar10}. \citet{bar10} confirmed the object as radial-mode pulsator and found $R=0.19\pm0.08$\,\rsol\, and $M=0.39^{+0.30}_{-0.13}$\,\msol using the Baade-Wesselink method. \citet{pie17} discovered a new class of radial mode pulsators with large amplitudes up to $300-400$\,mmag: the Blue-Large Amplitude pulsators (BLAPs). More recently a new class of radial mode hot subdwarf B type pulsators (high-gravity BLAPs) was discovered which show radial mode pulsation amplitudes up $150$\,mmag \citep{kup19}. Unlike most of the multi-periodic sdB pulsators, the BLAP and high-gravity BLAPS are more likely to be contracting low-mass helium pre-white dwarfs rather than compact He-core burning stars \citep{rom18, byr18, byr20}.

We discovered 12 new sdB pulsators which show only a single pulsation mode as part of this survey. We find amplitudes in the ZTF-$r$ band between $20$ and $73$ mmag which are similar to CS\,1246 \citep{bar10} but smaller than seen in the high-gravity BLAPs \citep{kup19}. Each system exhibits only a single frequency in the light curve (see Fig.\,\ref{fig:sdbvs} for the phase folded light curves). The pulsation periods range from 5.8\,minutes to 16 minutes. We performed a color correction for all objects using 3D extinction maps provided by \citet{gre19}\footnote{http://argonaut.skymaps.info/} and find very similar PanSTARRS colors between $g-r=-0.3$\,mag and $g-r=-0.45$\,mag. Using the reddening corrected $g$-band magnitudes and parallaxes from Gaia EDR3 \citep{gai16,gai20} we find absolute magnitudes in the range $\approx2.5-4.5$\,mag. Fig.\,\ref{fig:ztf_hr_sdbs} shows their position in a color-magnitude diagram color-coded with the pulsation period, which confirms that they are clustered in the region where hot subdwarf B stars are located. Table\,\ref{tab:sdbvs} presents an overview of the new pulsators. Remarkably, seven objects of our new sample show very similar pulsation periods around six minutes which is close to the pulsation period of CS\,1246, indicating that some of these new pulsators might be related to CS\,1246. If they are confirmed as radial-mode pulsators, like CS\,1246, future follow-up observations will measure their masses and radii. Phase resolved spectroscopy and multi-band photometry to fully characterize the objects is underway.



\section{Summary and Conclusions}\label{sec:sum}
We have presented an overview and first results of our dedicated high-cadence Galactic Plane survey for short period variable objects, carried out as part of ZTF during its first year of operation. We covered a total of 2990\,deg$^2$ between June 2018 and January 2019 in ZTF-$r$. Each field was observed continuously at a cadence of 40\,sec typically between $1.5$ and $3$ hours with some fields up $5-6$\,hours. We only selected fields with at least $80$ epochs and most fields have between $200$ and $400$. 

The survey covered a total of $\approx$230\,million stars to a limiting magnitude of 20.5\,mag in ZTF-$r$. Stellar densities range from 4 objects per arcmin$^{2}$ to $\approx$70 objects per arcmin$^{2}$. Four fields were selected for a detailed analysis. We calculate light curve statistics for each field and use the IQR to estimate the number of variable objects for each field. We find that $\approx1-2\,\%$ of objects show astrophysical variability in each field which leads to $\approx2-4$\,million expected variable objects in the ZTF Galactic Plane high-cadence Galactic Plane data. The largest contaminant are blended sources. For the highest density field this contamination rate is high, with up to $90\,\%$ of the sources found to be variable from the IQR. This shows that simple light curve statistics can reduce the number of non-variable sources substantially but is not sufficient to select a clean sample in high-density fields. Additional measures are required. In future, instead of using direct (non-difference) images, we will explore the use of forced photometry on difference images using input seed positions provided by external catalogs, e.g., {\em Gaia} EDR3. This will provide a more complete census of photometric variability in confused regions of the Galactic Plane. Forced photometry is expected to decrease the number of wrong magnitudes due to blending. Additionally, van \citet{roe21} has shown that machine learning classification can also substantially decrease the number of false positives. 

We have started to explore the data and present a progress report of recently published objects which were discovered by period searches using color-selected sub-samples. Some of these discoveries revealed new types of variable stars. In this work we also present a sample of 12 new single-mode hot subdwarf B-star pulsators with pulsation amplitudes between ZTF-$r=20 - 76$\,mmag and pulsation periods between $P=5.8 - 16$\,min. If they are confirmed as radial mode puslators future follow-up observations will measure their masses and radii.

Future work is in progress which applies machine learning techniques to classify large samples of individual objects \citep{cou20, roe21}. ZTF data up to December 2018  was released in data release 3 and the remaining data obtained in January 2019 was released in data release 4. ZTF has continued with high-cadence Galactic Plane observations in year 2 and 3. We settled on 140 epochs (1.5hrs) per field and continued to use ZTF-$r$ band. We aim to observe the entire Galactic Plane within $|b|<10$ and further out to high stellar density regions. We continue to analyse the data using period-finding algorithms and light curve statistics, but are also exploring the use of neural networks to identify and classify variable stars.


Looking ahead, the Vera Rubin Observatory's Legacy Survey of Space and Time (LSST) will observe to fainter magnitudes and with an expected precision of $\sim$\,5 mmag is expected to find more variable sources \citep{hub06}. Given the significantly larger number of sources in the LSST the Galactic Plane observations of LSST are expected to observe unique sources which are too faint for ZTF but the high stellar density will challenge data analysis and the significantly higher number of sources will require more computational power to search all the objects.

\section*{Acknowledgements}
We thank the anonymous referee for their helpful comments and review of the manuscript.

Based on observations obtained with the Samuel Oschin Telescope 48-inch and the 60-inch Telescope at the Palomar Observatory as part of the Zwicky Transient Facility project. ZTF is supported by the National Science Foundation under Grant No. AST-1440341 and a collaboration including Caltech, IPAC, the Weizmann Institute for Science, the Oskar Klein Center at Stockholm University, the University of Maryland, the University of Washington, Deutsches Elektronen-Synchrotron and Humboldt University, Los Alamos National Laboratories, the TANGO Consortium of Taiwan, the University of Wisconsin at Milwaukee, and Lawrence Berkeley National Laboratories. Operations are conducted by COO, IPAC, and UW.

This research was funded by the Gordon and Betty Moore Foundation through Grant GBMF5076. This research was supported in part by the National Science Foundation under Grant No. NSF PHY-1748958. 

M.W.C. acknowledges support from the National Science Foundation with grant number PHY-2010970.

This work has made use of data from the European Space Agency (ESA) mission {\it Gaia} (\url{https://www.cosmos.esa.int/gaia}), processed by the {\it Gaia} Data Processing and Analysis Consortium (DPAC,
\url{https://www.cosmos.esa.int/web/gaia/dpac/consortium}). Funding for the DPAC
has been provided by national institutions, in particular the institutions
participating in the {\it Gaia} Multilateral Agreement.

\section*{Data Availability}

All the observational data used in this paper are publicly available at the IRSA database and can be accessed here: \url{https://www.ztf.caltech.edu/page/dr4}.




\bibliographystyle{mnras}
\bibliography{refs,refs_1508} 




\appendix

\section{Fields observed as part of the high-cadence Galactic Plane observations in ZTF year-1}

\begin{table*}
	\centering
	\caption{Fields observed as part of the June/July campaign. Median values are given for limiting magnitude and seeing}
	\label{tab:fields_summer}
	\begin{tabular}{lcrrrrrrrrr}
	\hline
Date  &  ZTF   & RA &  Dec &  Gal Lat &  Gal Long &   epochs & limit  &  seeing & duration  & fraction  \\
UTC & FieldID &  (deg)  & (deg) & (deg) & (deg)  &  & (mag) & (arcsec) &   (hrs)  &  successful \\
\hline \smallskip
06-16-18 & 284 &  279.8366  &  $-$24.2500 &   $-$8.9125 &    10.2331  &     23  & 20.72 & 1.98 & 0.23 & 0.98\\ \smallskip
06-17-18 & 284 &  279.8366 &  $-$24.2500 &   $-$8.9125 &    10.2331  &     113  & 20.58 & 2.26  & 1.23 & 0.96 \\ \smallskip
06-18-18 & 284 &  279.8366 &  $-$24.2500 &   $-$8.9125 &    10.2331  &     107  & 20.39 & 2.51  & 1.16 & 0.99 \\ \smallskip
06-20-18 & 330 &  255.5263 &  $-$17.0500 &   14.2859 &     4.7848  &      115  & 20.32 & 2.51  & 1.25 &  1.00\\ \smallskip
06-21-18 &   330 &  255.5263 &  $-$17.0500 &   14.2859 &   4.7848     &   115  & 20.36 & 2.21   & 1.25 &  1.00 \\ \smallskip
06-22-18 &   385 &  287.6598 &   $-$9.8500  &  $-$9.2762  &   26.6514    &   115  & 20.54 & 2.34  & 1.38 & 0.73 \\  \smallskip
06-23-18 &   385 &  287.6598 &   $-$9.8500  &  $-$9.2762  &   26.6514    &    115  & 20.47 & 2.15  & 1.25  & 0.81\\ \smallskip
06-24-18 &   385 &  287.6598 &   $-$9.8500  &  $-$9.2762  &   26.6514    &    115  & 20.07 & 2.42  & 1.25 & 0.88 \\ \smallskip
06-25-18 &   766 &  292.6968 &   47.7500 &   13.4390 &    80.1691  &      113  & 20.75 & 1.59  & 1.27 & 1.00 \\ \smallskip
06-26-18 &   766 &  292.6968 &   47.7500 &   13.4390 &    80.1691   &     113  & 20.47 & 1.82  & 1.27 & 1.00  \\ \smallskip
06-27-18 &   766 &  292.6968 &   47.7500 &   13.4390  &   80.1691   &     111  & 20.32 & 1.90  & 1.28 &  1.00  \\ \smallskip
06-28-18 &   727 &  287.0350 &   40.5500 &   14.0583  &   71.6873   &     115  & 20.05 & 2.33  & 1.25 &  1.00  \\ \smallskip
06-29-18 &   727 &  287.0350 &   40.5500 &   14.0583  &   71.6873   &     137  & 19.74 & 3.35  & 1.49  &  1.00    \\ \smallskip
06-30-18 &   684 &  282.5628 &   33.3500 &   14.4152  &   63.4051   &   133  & 20.34 & 2.13  & 1.45  & 1.00 \\ \smallskip
07-01-18 &    684 &  282.5628 &   33.3500 &   14.4152  & 63.4051  &   133  & 20.33 & 1.84  & 1.45  & 1.00 \\ \smallskip
07-03-18 &    280 &  257.8188 &  $-$24.2500 &    8.3828  &    0.0438   &     133  & 20.32 & 2.79  & 1.45  & 0.74 \\  \smallskip
07-04-18 &    280 &  257.8188 &  $-$24.2500 &    8.3828  &    0.0438   &     132  & 20.37 & 2.22  & 1.44  & 0.67 \\ \smallskip
07-05-18 &    331 &  262.5569 &  $-$17.0500 &    8.6904  &    8.5864   &   134  & 20.41 & 2.33 &  1.46  & 0.49 \\   \smallskip
07-06-18  &   331 &  262.5569 &  $-$17.0500 &    8.6904 &     8.5864   &     133  & 20.04 & 2.85  & 1.45  & 0.68 \\ \smallskip
07-07-18  &   331 &  262.5569 &  $-$17.0500 &    8.6904 &     8.5864   &     133  & 20.33 & 2.44  & 1.45 & 0.72 \\ \smallskip
07-08-18  &   282 &  272.4435 &  $-$24.2500 &   $-$2.9315 &     7.1078   &     133  & 20.26 & 2.31  &  1.45  & 0.43 \\ \smallskip
07-09-18  &   282 &  272.4435  &  $-$24.2500 &   $-$2.9315 &     7.1078   &     135  & 20.09 & 2.52  &  1.47  & 0.44 \\   \smallskip
07-11-18 &   283 &  279.8366 &  $-$24.2500 &   $-$8.9125 &    10.2331 &      116  & 20.14 & 2.40   & 1.26  &  0.40 \\ \smallskip
07-12-18 &   283 &  279.8366 &  $-$24.2500 &   $-$8.9125 &    10.2331    &    132  & 20.26 & 2.25 &  1.44 & 0.47 \\ \smallskip
07-13-18 &   283 &  279.8366 &  $-$24.2500 &   $-$8.9125 &    10.2331    &    133  & 20.27 & 2.30  & 1.45 & 0.46 \\ \smallskip
07-14-18 &   333 &  276.6440 &  $-$17.0500 &   $-$2.9817 &    15.3198    &   133   & 20.01 & 2.22 &  1.45  & 0.23 \\ \smallskip
07-15-18 &   333 &  276.6440 &  $-$17.0500 &   $-$2.9817 &    15.3198    &   133  & 19.97 & 2.34 &  1.45  & 0.32 \\ \smallskip
07-16-18 &   333 &  276.6440 &  $-$17.0500 &   $-$2.9817 &    15.3198    &   136  & 19.86 & 2.17  &  1.48 &  0.54 \\ \smallskip
07-17-18 &   384 &  280.6106 &   $-$9.8500  &  $-$3.0783  &   23.4976  &     134  & 20.22 & 2.23  &  1.48 & 0.40  \\ \smallskip
07-18-18 &   384 &  280.6106 &   $-$9.8500  &  $-$3.0783  &   23.4976  &     125  & 20.52 & 2.09  & 1.36  & 0.36 \\ \smallskip
07-19-18 &   384 &  280.6106 &   $-$9.8500  &  $-$3.0783  &   23.4976  &     136  & 20.26 & 2.16   & 1.48  &  0.33 \\ \smallskip
07-20-18 &   334 &  283.7259 &  $-$17.0500 &   $-$9.0209 &    18.3982 &      135  & 20.01 & 2.63  &  1.47 & 0.72 \\ \smallskip
07-21-18 &   334 &  283.7259 &  $-$17.0500  &  $-$9.0209  &   18.3982 &      136  & 20.14 & 1.97  &  1.48 &  1.00 \\ \smallskip
07-22-18 &   334 &  283.7259 &  $-$17.0500  &  $-$9.0209  &   18.3982    &   136  & 20.10 & 2.02 & 1.48  &  1.00 \\ \smallskip
07-23-18 &   683 &  274.7095 &   33.3500  &  20.5066 &    60.8553   &     137  & 20.63 & 1.73  &   1.49  & 1.00 \\ \smallskip
07-24-18 &   683 &  274.7095 &   33.3500  &  20.5066 &    60.8553   &    134  & 20.39 & 1.80   & 1.46  &  1.00 \\ \smallskip
07-25-18 &   726 &  278.4181 &   40.5500  &  20.1368 &    69.1476   &     133  & 20.10 & 1.86  &  1.45  & 1.00 \\ \smallskip
07-26-18 &   726 &  278.4181 &   40.5500 &   20.1368 &    69.1476   &     133  & 20.03 & 1.85 &  1.45  &  1.00 \\ \smallskip
07-27-18 &   765 &  282.9221 &   47.7500 &   19.4902 &    77.5052   &     133  & 20.20 & 1.72  &  1.45  &  1.00 \\ \smallskip
07-28-18 &   765 &  282.9221 &   47.7500 &   19.4902 &    77.5052   &     133  & 20.06 & 1.70  &  1.45    &  1.00 \\ \smallskip
07-29-18 &   330 &  255.5263 &  $-$17.0500 &   14.2859 &     4.7848    &       69  & 19.91 & 1.75  & 1.19 & 1.00 \\ \smallskip
07-31-18 &  282 & 272.4435 & $-$24.2500 &  $-$2.9315  &    7.1078         &  100   & 19.71 & 2.12   &  1.08 & 0.80 \\ 
		\hline
	\end{tabular}
\end{table*}

\begin{table*}
	\centering
	\caption{Fields observed as part of the August campaign. Median values are given for limiting magnitude and seeing}
	\label{tab:fields_fall}
	\begin{tabular}{lcrrrrrrrrr}
	\hline
Date  &  ZTF   & RA &  Dec &  Gal Lat &  Gal Long &   epochs & limit  &  seeing & duration & fraction  \\
UTC & FieldID &  (deg)  & (deg) & (deg) & (deg)  &  & (mag) & (arcsec) & (hrs)  & successful \\
\hline
08-03-2018 &    538  &    284.8764  &   11.7500   &   3.0889   &   44.6385  &    118  &  20.84 & 1.87 & 1.28  & 1.00 \\ \smallskip
08-03-2018 &    539  &    277.9013  &   11.7500   &   9.1851   &   41.5422  &    42   &  20.66  & 1.97  & 0.45 & 1.00  \\ 
08-04-2018 &    538  &    284.8764  &   11.7500   &   3.0889   &   44.6385  &    118  &  20.93  &  1.87  & 2.59  & 1.00 \\  \smallskip
08-04-2018 &    539  &    277.9013  &   11.7500   &   9.1851   &   41.5422  &    118  & 20.76 &  2.19  & 2.59  &  1.00 \\
08-07-2018 &    638  &   285.7714   &  26.1500  &   2.9069   &   61.0817  &    118   & 21.09  & 1.89 &  2.67  &  1.00 \\  \smallskip
08-07-2018 &    639  &  293.2531  &  26.1500    &    $-$2.8742  &   64.5379  &    118   &  20.97  & 2.08  &  2.67  &  1.00 \\
08-08-2018 &    638  &  285.7714   &  26.1500    &     2.9069   &   61.0817  &    118   &  21.03 &  1.84  & 2.67  &  1.00 \\  \smallskip
08-08-2018 &    639  &  293.2531  &  26.1500  &    $-$2.8742  &   64.5379  &    118   & 20.90  &  2.03   &  2.67  &  1.00 \\
08-10-2018 &   488  &   288.1194   &   4.5500   &  $-$3.0953    &  39.7486  &    118   & 20.55  & 2.01 &  2.59 &  1.00 \\  \smallskip
08-10-2018 &   489  &  295.1303    &  4.5500    &   $-$9.2520   &   43.0768  &    118   &  20.42  & 2.14  &  2.59   & 1.00 \\
08-11-2018 &   540  &    291.9311  &   11.7500   &  $-$3.0068    &  47.8872   &    54   &  20.52 &  2.00 & 2.31  & 1.00  \\  \smallskip
08-11-2018 &   541  &   298.9153   &  11.7500   &   $-$8.9492   &   51.2901  &    55   & 20.54  &  2.04 & 1.26  & 1.00  \\
08-12-2018 &   540  &   291.9311  &   11.7500   &  $-$3.0068    &  47.8872 &    117   &  20.67 &  2.12 &  2.57 & 1.00 \\  \smallskip
08-12-2018 &   541  &   298.9153   &  11.7500   &   $-$8.9492   &   51.2901  &    116   & 20.57  &  2.22 & 2.55  &  1.00 \\
08-13-2018 &   591  &   296.1269   &   18.9500  &  $-$2.9775    &   56.1558  &    118   &  20.80 &  2.09 &  2.61 &  1.00 \\  \smallskip
08-13-2018 &   592  &   303.3767   &    18.9500  &   $-$8.8128   &  59.7765  &    118   & 20.85  &  2.01 &  2.61  & 1.00  \\
08-14-2018 &   591  &    296.1269   &   18.9500  &  $-$2.9775    &   56.1558    &    118   &  20.92 &  1.87 &  2.63 & 1.00 \\  \smallskip
08-14-2018 &   592  &   303.3767   &    18.9500  &   $-$8.8128   &  59.7765  &    118   & 20.95  &  1.95 &   2.63 & 1.00  \\
08-15-2018 &   685  &   290.4160   &   33.3500   &   8.4910   &  66.2776   &    117   &  20.92 &  1.95 &  2.77 & 1.00 \\ \smallskip
08-15-2018 &   686  &   298.2693   &    33.3500  &   2.7706   &   69.5025  &    117   &  20.77  &  2.09 & 2.77 & 1.00 \\
08-16-2018 &   685  &   290.4160   &   33.3500   &   8.4910   &  66.2776  &    112   &  20.93 &  1.81 & 2.81  & 0.85 \\ \smallskip
08-16-2018 &   686  &   298.2693   &    33.3500  &   2.7706   &   69.5025  &    111   & 20.81  &  1.95 &  2.79 &  0.86  \\
08-17-2018 &   436  &   284.3894   &  $-$2.65000    &  $-$3.1176    &  31.6292   &    97    &  20.37 &  2.19 & 2.79  & 0.98 \\ \smallskip
08-17-2018 &   437  &   291.4182   &  $-$2.65000  &   $-$9.3570   &  34.8701   &    94    &  20.47  &  2.12 &  2.79  & 1.00 \\
08-18-2018 &   436  &  284.3894   &  $-$2.65000    &  $-$3.1176    &  31.6292    &    118   &  20.51 &   2.16 &  2.57 &  0.97  \\ \smallskip
08-18-2018 &   437  &   291.4182   &  $-$2.65000  &   $-$9.3570   &  34.8701   &    118   &  20.61  &  2.04 & 2.57 & 1.00 \\
		\hline
	\end{tabular}
\end{table*}

\begin{table*}
	\centering
	\caption{Fields observed as part of the winter campaign. Median values are given for limiting magnitude and seeing}
	\label{tab:fields_winter}
	\begin{tabular}{lcrrrrrrrrr}
	\hline
Date  &  ZTF   & RA &  Dec &  Gal Lat &  Gal Long &   epochs & limit  &  seeing  & duration  & fraction  \\
UTC & FieldID &  (deg)  & (deg) & (deg) & (deg)  &  & (mag) & (arcsec) &  (hrs) & successful \\
\hline
11-15-2018 &   728  &   295.6518  & 40.5500 &  8.2113  &  74.6857   &   148   &  19.93 &   2.81 &  3.29  & 1.00 \\ \smallskip
11-15-2018 &   729  &   304.2686  & 40.5500 &  2.6520  &  78.1565   &   148   &  20.04  &  2.74 &  3.29  & 1.00 \\
11-19-2018 &   799  &   291.0910  & 54.9500 &  17.3604   &  86.5193  &  145   &  20.16 &   1.72 &  3.32  & 0.92 \\ \smallskip
11-19-2018 &   800  &   302.1884  & 54.9500 &  11.7051   &  89.5415  &  146   &  20.08  &  1.74 &  3.90  & 0.90 \\
11-26-2018 &   767  &   302.4714  & 47.7500 &   7.7191   &  83.4543  &  128   &  20.30  &  1.81 &  2.88  & 1.00 \\ \smallskip
11-26-2018 &   768  &   312.2461  & 47.7500 &   2.4194   &  87.3502  &  128   &  20.42  &  1.81 &  2.88  & 1.00 \\
11-27-2018 &   688  &   313.8717  & 33.3500 & $-$7.8072  &  77.1136  &  156   &  20.41  &  2.14 &  4.10  & 0.99 \\ \smallskip
11-27-2018 &   689  &   321.6554  & 33.3500 & $-$12.5790 &  81.6002  &  158   &  20.48  &  2.20 &  4.20  & 1.00 \\
11-28-2018 &   688  &   313.8717  & 33.3500 & $-$7.8072  &  77.1136  &  142   &  19.66  &  1.91 &  3.75  & 0.80 \\ \smallskip
11-28-2018 &   689  &   321.6554  & 33.3500 & $-$12.5790 &  81.6002  &  142   &  19.78  &  1.87 &  3.76  & 0.86 \\
11-29-2018 &   767  &   302.4714  & 47.7500 &   7.7191   &  83.4543  &  127   &  20.62  &  1.81 &  2.86  & 1.00 \\ \smallskip
11-29-2018 &   768  &   312.2461  & 47.7500 &   2.4194   &  87.3502  &  133   &  20.67  &  1.80 &  2.86  & 1.00 \\
12-03-2018 &   730  &   312.8218  & 40.5500 & $-$2.5179  &  82.0919  &  59    &  20.62  &  1.81 &  1.30  & 1.00 \\ 
12-03-2018 &   731  &   321.3116  & 40.5500 & $-$7.2339  &  86.5070  &  109   &  20.67  &  1.80 &  1.84  & 0.98 \\ \smallskip
12-03-2018 &   259  &   104.6634 & $-$24.2500  &  $-$8.9125   &   235.7669  &  180   &  20.67  &  1.80 &  1.96  & 0.95 \\
12-04-2018 &   411  &   107.1700 & $-$2.6500   &  3.1405  &  217.6228  &  275   &  20.29  &  2.25 &  6.02  & 0.85 \\ \smallskip
12-04-2018 &   412  &   114.2600 & $-$2.6500   &  9.3964  &  220.9648  &  276   &  20.30  &  2.30 &  6.02  & 0.86 \\
12-05-2018 &   411  &   107.1700 & $-$2.6500   &  3.1405  &  217.6228  &  275   &  20.36  &  2.41 &  6.04  & 1.00 \\ \smallskip
12-05-2018 &   412  &   114.2600 & $-$2.6500   &  9.3964    & 220.9648  &  275   &  20.39  &  2.46 &  6.01  & 1.00 \\
12-09-2018 &   770  &   332.1016 &   47.7500   & $-$6.6404  & 97.1225   &  134   &  20.24  &  2.35 &  3.97  & 1.00 \\
12-09-2018 &   771  &   342.1387 &   47.7500   & $-$10.1214 & 102.9618  &  132   &  20.42  &  2.26 &  3.95  & 1.00 \\ \smallskip
12-09-2018 &   259  &   104.6634 & $-$24.2500  &  $-$8.9125   &  235.7669   &  111   &  19.77  &  2.19 &  1.26  & 0.95 \\
12-12-2018 &   308  &   100.7741 & $-$17.0500  &  $-$9.0209   &  227.6018   &  140   &  20.32  &  2.53 &  3.05  & 0.97 \\ \smallskip
12-12-2018 &   309  &   107.8560 & $-$17.0500  &  $-$2.9817   &  230.6802   &  140   &  20.26  &  2.66 &  3.05  & 1.00 \\
12-13-2018 &   308  &   100.7741 & $-$17.0500  &  $-$9.0209   &  227.6018   &  78    &  19.96  &  3.16 &  1.69  & 1.00 \\ 
12-13-2018 &   309  &   107.8560 & $-$17.0500  &  $-$2.9817   &  230.6802   &  78    &  19.99  &  3.14 &  1.69  & 1.00 \\ \smallskip
12-13-2018 &   769  &   322.0645 &    47.7500  &  $-$2.3827   &   91.8751   &  269   &  19.83  &  3.35 &  2.94  & 1.00 \\ \smallskip
12-14-2018 &   769  &   322.0645 &    47.7500  &  $-$2.3827   &   91.8751   &  270   &  20.06  &  2.36 &  2.95  & 0.92 \\ 
12-16-2018 &   803  &   335.1841 &    54.9500  &  $-$1.7248   &  102.7388   &  345   &  20.21  &  2.11 &  3.77  & 1.00 \\ \smallskip
12-16-2018 &   257  &    89.8772 & $-$24.2500  & $-$21.2513   &  230.1444   &  198   &  20.42  &  2.02 &  2.16  & 1.00 \\ 
12-17-2018 &   512  &    92.5688 &    11.7500  &  $-$3.0068   &  198.1128   &  201   &  19.85  &  1.93 &  6.14  & 0.87 \\ \smallskip
12-17-2018 &   513  &    99.6235 &    11.7500  &     3.0889.  &  201.3615   &  201   &  19.90  &  1.94 &  6.14  & 0.87 \\ \smallskip
12-19-2018 &   257  &    89.8772 & $-$24.2500  & $-$21.2513   &  230.1444   &  189   &  19.74  &  2.50 &  2.06  & 1.00 \\ \smallskip
12-20-2018 &   257  &    89.8772 & $-$24.2500  & $-$21.2513   &  230.1444   &  188   &  19.60  &  2.42 &  2.05  & 1.00 \\ 
12-22-2018 &   807  &    20.0000 &    62.1500  & $-$0.2329    &  126.6119   &  271   &  19.91  &  1.82 &  2.96  & 1.00 \\ \smallskip
12-22-2018 &   310  &   114.9123 & $-$17.0500  &    2.9306    &  233.9298   &  279   &  19.30  &  2.36 &  3.05  & 0.91 \\
12-23-2018 &   807  &    20.0000 &    62.1500  & $-$0.2329    &  126.6119   &  270   &  19.82  &  2.11 &  2.95  & 1.00 \\ \smallskip
12-23-2018 &   310  &   114.9123 & $-$17.0500  &    2.9306    &  233.9298   &  136   &  19.41  &  2.18 &  1.48  & 0.94 \\
12-24-2018 &   804  &   345.7743 &    54.9500  & $-$4.5684    &  108.1171   &  153   &  20.06  &  1.75 &  3.49  & 0.94 \\ 
12-24-2018 &   805  &   356.3646 &    54.9500  & $-$6.5258    &  113.8966   &  153   &  20.17  &  1.71 &  3.49  & 0.95 \\ \smallskip
12-24-2018 &   260  &   112.0564 & $-$24.2500  & $-$2.9315    &  238.8922   &  189   &  19.52  &  2.10 &  2.06  & 0.81 \\ 
12-28-2018 &   804  &   345.7743 &    54.9500  & $-$4.5684    &  108.1171   &  58    &  19.60  &  3.87 &  1.31  & 0.82 \\ 
12-28-2018 &   805  &   356.3646 &    54.9500  & $-$6.5258    &  113.8966   &  87    &  19.83  &  3.57 &  1.28  & 0.88 \\ \smallskip
12-28-2018 &   260  &   112.0564 & $-$24.2500  & $-$2.9315    &  238.8922   &  196   &  19.44  &  3.83 &  2.14  & 0.94 \\ 
12-29-2018 &   806  &     6.6666 &    62.1500  & $-$0.3373    &  120.3949   &  87    &  19.86  &  3.84 &  0.84  & 1.00 \\ \smallskip
12-29-2018 &   260  &   112.0564 & $-$24.2500  & $-$2.9315    &  238.8922   &  196   &  19.75  &  3.25 &  2.14  & 1.00 \\
12-30-2018 &   462  &   110.1486 &     4.5500  &    9.0982    &  212.5713   &  275   &  20.09  &  2.91 &  6.01  & 0.98 \\
12-30-2018 &   463  &   116.9904 &     4.5500  &   15.1705    &  215.7493   &  274   &  20.09  &  2.96 &  5.99  & 0.98  \\
\hline
\end{tabular}
\end{table*}

\begin{table*}
	\centering
	\begin{tabular}{lcrrrrrrrrr}
	\multicolumn{1}{}{} {\bf Table A3} continued. \\
		\hline
Date  &  ZTF   & RA &  Dec &  Gal Lat &  Gal Long &   epochs & limit  &  seeing & duration & fraction  \\
UTC & FieldID &  (deg)  & (deg) & (deg) & (deg)  &  & (mag) & (arcsec) &  (hrs)  & successful \\
\hline \smallskip
01-03-2019 &   261  &  119.3957  & $-$24.2500  &     2.8444  &   242.2711    &   172   &  19.55  & 4.05 &  1.87  & 0.93 \\ 
01-04-2019 &   261  &  119.3957  & $-$24.2500  &     2.8444  &   242.2711    &   171   &  20.00  & 2.83 &  1.83  & 1.00 \\ \smallskip
01-04-2019 &   258  &   97.2703  & $-$24.2500  & $-$15.0272  &   232.8673    &   163   &  20.22  & 2.60 &  1.78  & 1.00 \\ \smallskip
01-08-2019 &   360  &  110.9488  &  $-$9.8500  &     3.0707  &   225.7543    &   333   &  19.91  & 3.57 &  3.64  & 0.91 \\
01-10-2019 &   262  &  126.6811  & $-$24.2500  &     8.3828  &   245.9562    &   172   &  20.32  & 2.29 &  1.87  & 0.95 \\ \smallskip
01-10-2019 &   260  &  112.0564  & $-$24.2500  &  $-$2.9315  &   238.8922    &   159   &  20.13  & 2.58 &  1.73  & 1.00 \\
01-11-2019 &   311  &  121.9430  & $-$17.0500  &     8.6904  &   237.4136    &   275   &  20.58  & 2.32 &  3.00  & 1.00 \\
		\hline

	\end{tabular}
\end{table*}

\bsp	
\label{lastpage}
\end{document}